\documentclass[pra,twocolumn,superscriptaddress,nobibnotes,
notitlepage,nofootinbib,floatfix]{revtex4-1}

\usepackage{soul}
\usepackage{graphicx}
\usepackage{amssymb,amsmath,color,amsfonts, amsthm}
\usepackage{hyperref}
\usepackage{ragged2e}
\usepackage[capitalise]{cleveref}
\usepackage[normalem]{ulem}

\hypersetup{
	colorlinks,
	linkcolor={blue},
	citecolor={blue},
	urlcolor={blue}
}

\usepackage{subfigure}

\theoremstyle{plain}

\newcommand{\ket}[1]{|#1\rangle}

\newcommand{\mc}{\ensuremath{\mathcal}}
\newcommand{\mr}{\ensuremath{\mathrm}}

\newcommand{\remove}[1]{\ifmmode\text{\sout{\ensuremath{\color{red}#1}}}\else{\color{red}\sout{#1}}\fi}

\newcommand{\fps}[1]{|#1\,\%}

\begin{document}

\title{Entangling logical qubits with lattice surgery}

\date{\today}

\author{Alexander Erhard}
\thanks{These authors contributed equally to this work. Contact: alexander.erhard@uibk.ac.at, hendrik.poulsen-nautrup@uibk.ac.at}
\affiliation{Institute for Experimental Physics, University of Innsbruck, 6020 Innsbruck, Austria}

\author{Hendrik Poulsen Nautrup}
\thanks{These authors contributed equally to this work. Contact: alexander.erhard@uibk.ac.at, hendrik.poulsen-nautrup@uibk.ac.at}
\affiliation{Institute for Theoretical Physics, University of Innsbruck, 6020 Innsbruck, Austria}

\author{Michael Meth}
\affiliation{Institute for Experimental Physics, University of Innsbruck, 6020 Innsbruck, Austria}

\author{Lukas Postler}
\affiliation{Institute for Experimental Physics, University of Innsbruck, 6020 Innsbruck, Austria}

\author{Roman Stricker}
\affiliation{Institute for Experimental Physics, University of Innsbruck, 6020 Innsbruck, Austria}

\author{Martin Ringbauer}
\affiliation{Institute for Experimental Physics, University of Innsbruck, 6020 Innsbruck, Austria}

\author{Philipp Schindler}
\affiliation{Institute for Experimental Physics, University of Innsbruck, 6020 Innsbruck, Austria}

\author{Hans J. Briegel}
\affiliation{Institute for Theoretical Physics, University of Innsbruck, 6020 Innsbruck, Austria}
\affiliation{Fachbereich Philosophie, Universit{\"a}t Konstanz, Fach 17, 78457 Konstanz, Germany}

\author{Rainer Blatt}
\affiliation{Institute for Experimental Physics, University of Innsbruck, 6020 Innsbruck, Austria}
\affiliation{Institute for Quantum Optics and Quantum Information of the Austrian Academy of Sciences, 6020 Innsbruck, Austria}

\author{Nicolai Friis}
\affiliation{Institute for Quantum Optics and Quantum Information - IQOQI Vienna, Austrian Academy of Sciences, 1090 Vienna, Austria}
\affiliation{Institute for Theoretical Physics, University of Innsbruck, 6020 Innsbruck, Austria}

\author{Thomas Monz}
\affiliation{Institute for Experimental Physics, University of Innsbruck, 6020 Innsbruck, Austria}
\affiliation{Alpine Quantum Technologies GmbH, 6020 Innsbruck, Austria}

\begin{abstract}
Future quantum computers will require quantum error correction for faithful operation. The correction capabilities come with an overhead for performing fault-tolerant logical operations on the encoded qubits. One of the most resource efficient ways to implement logical operations is lattice surgery, where groups of physical qubits, arranged on lattices, can be merged and split to realize entangling gates and teleport logical information. Here, we report on the experimental realization of lattice surgery between two topologically encoded qubits in a 10-qubit ion trap quantum information processor. In particular, we demonstrate entanglement between two logical qubits and we implement logical state teleportation.
\end{abstract}

\maketitle

\vspace*{-4.0mm}
The development of quantum computing architectures from early designs and current noisy intermediate-scale quantum (NISQ) devices~\cite{Preskill1997} to full-fledged quantum computers hinges on achieving fault-tolerance using quantum error correction (QEC)~\cite{Preskill1997, NielsenChuang2000}. The basis of QEC is storing and manipulating quantum information using logical qubits. A number of experiments have demonstrated significant technological progress towards QEC~\cite{CoryEtAl1998, KnillLaflammeMartinezNegrevergne2001, ChiaveriniEtal2004, BoulantViolaFortunatoCory2005, ZhangGangloffMoussaLaflamme2011, WoottonLoss2018}, including the creation of non-trivial QEC codes~\cite{BellHerreraMartiTameMarkhamWadsworthRarity2014, TakitaEtAl2017}, error detection~\cite{Kelly-Martinis2015, LinkeEtAl2017, Andersen2019}, correction of errors~\cite{AokiEtAl2009, ReedEtal2012, WaldherrEtAl2014, OfekEtAl2016} and qubit loss~\cite{StrickerEtal2020}, and operations on single~\cite{ZhangLaflammeSuter2012, NiggEtAl2014, Barends-Martinis2014, HeeresEtAl2017, Gong-Pan2019, HuEtAl2019} and on two logical qubits in non-topological codes~\cite{ChouEtal2018, HarperFlammia2019}.

The most promising road towards QEC is offered by topological codes, such as the surface code~\cite{Kitaev2003,DennisKitaevLandahlPreskill2002,FowlerMariantoniMartinisCleland2012}, which require only short-range interactions in 2D architectures. Nevertheless, the implementation of encoded operations remains a major challenge. Performing arbitrary logical operations requires costly techniques, including transversal gates~\cite{GottesmanPhD1997}, teleported gates~\cite{Gottesman1999}, and magic state distillation~\cite{BravyiKitaev2005}. Recent theoretical advances led to the development of lattice surgery (LS)~\cite{HorsmanFowlerDevittVanMeter2012,PoulsenNautrupFriisBriegel2017, GutierrezMuellerBermudez2019}, promising to reduce this complexity~\cite{Litinski2019}, while maintaining 2D layouts. In LS, the QEC code itself is altered by merging and splitting initially separate encodings, rather than operating on all physical qubits. Such modifications can be used to efficiently manipulate logical qubits, or to adapt the robustness to different noise processes~\cite{PoulsenNautrupDelfosseDunjkoBriegelFriis2019}. LS further enables entanglement generation between logical qubits and can be complemented with measurement-based protocols~\cite{raussendorf2001one,raussendorf2007fault,lanyon2013measurement} for logical state teleportation and manipulation~\cite{PoulsenNautrupFriisBriegel2017}. Here, we report the experimental implementation of LS using $10$ trapped ions to entangle two logical qubits encoded in two $4$-qubit surface codes~\cite{FowlerMariantoniMartinisCleland2012}.

%
\begin{figure*}[ht!]
\centering
  \includegraphics[width=0.95\textwidth]{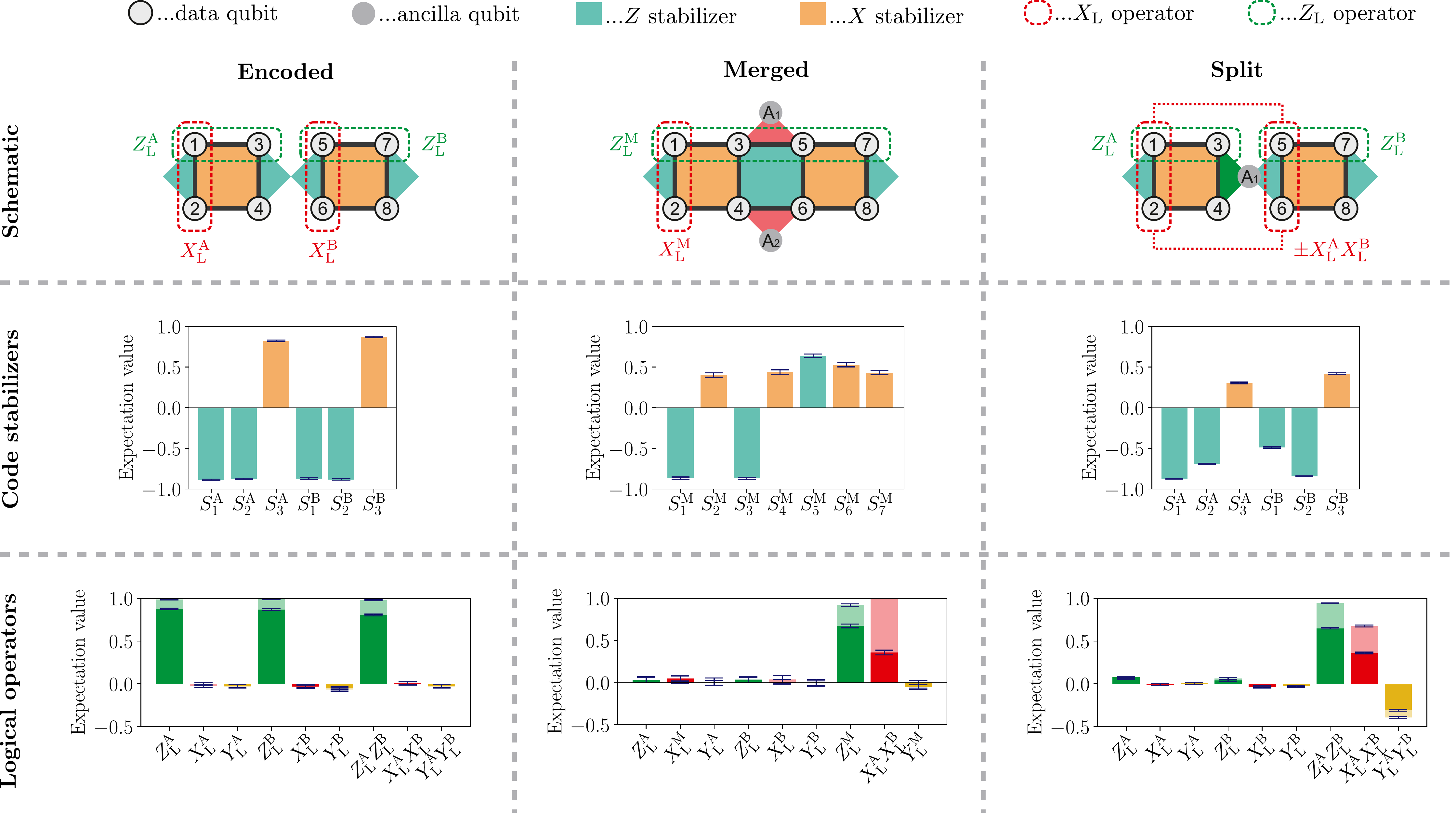}
\caption{\textbf{Experimental surface code lattice surgery.}
Experimental results and schematics for LS between $Z$-type (rough) boundaries implementing a logical joint measurement $M^{\pm}_\mr{XX}=\mathbb{I}\pm X_\mr{L}^\mr{A}X_\mr{L}^\mr{B}$ to generate a logical Bell state. We use the error detection capabilities of the code and post-select measurements with valid code stabilizers, which are presented in light colored bars.
\textbf{Encoded:} Two surface codes defined on $2\times 2$ lattices with average code stabilizer values of $\langle |S_i|\rangle=0.868(4)$ (error is calculated from individual stabilizer errors) where $X$-stabilizers and $Z$-stabilizers in~\cref{eq:sc_stabilizer} are associated with orange and aquamarine faces, respectively. We observe (raw$|$post selected) state fidelities $\mc{F}(\ket{0_\mr{L}^\mr{A}})=93.8(4)\fps{99.3(2)}$ and ${\mc{F}(\ket{0_\mr{L}^\mr{B}})=93.4(5)\fps{99.4(2)}}$ for the encodings $\ket{0^\mr{A}_\mr{L}}$, $\ket{0^\mr{B}_\mr{L}}$, respectively.
Logical operators are products of Pauli operators connecting opposite boundaries (see \cref{eq:sc_logical}).
\textbf{Merged:} Stabilizers along the boundaries are measured (red) using ancillas $A_1,A_2$ such that ${S_6^\mr{M}S_7^\mr{M}=X_\mr{L}^\mr{A}X_\mr{L}^\mr{B}}$. The merged code~\cref{eq:sc_merged_rough} encodes a single logical qubit $\ket{0_\mr{L}^\mr{M}}$ corresponding to the logical operator $Z_\mr{L}^\mr{M}=Z_\mr{L}^\mr{A}Z_\mr{L}^\mr{B}$ in~\cref{eq:sc_merged_logic_rough}. We observe average stabilizer values and logical state fidelities of $\langle|S_i|\rangle=0.669(8)$, $\mc{F}(\ket{0_\mr{L}^\mr{M}})=86.4(1.0)\fps{97.9(5)}$, respectively.
\textbf{Split:} In order to split the merged code while preserving the eigenstate of $X_\mr{L}^\mr{A}X_\mr{L}^\mr{B}$, one boundary stabilizer of the original code is measured (green) reusing ancilla $A_1$. In this way, we recover the original codes with average stabilizer values of $\langle|S_i|\rangle=0.603(3)$ which are now in a logical Bell state $\ket{\phi_\mr{L}^+}$ with fidelity $\mc{F}(\ket{\phi_\mr{L}^+})=58.0(1.6)\fps{75.3(1.6)}$.}
\label{fig:sc_ls}
\end{figure*}

\noindent \textbf{Surface code}.
One of the most prominent examples of a QEC code is the surface code~\cite{Kitaev2003,DennisKitaevLandahlPreskill2002,FowlerMariantoniMartinisCleland2012,raussendorf2007fault} which has error thresholds of up to $1\%$~\cite{WangFowlerHollenberg2011}. The surface code has a simple description within the stabilizer formalism~\cite{GottesmanPhD1997}, as we discuss in the following (see Appendix~\ref{app:intro_qec} for more details).

Here, we consider the minimal instance of a surface code \textemdash\ a 4-qubit code encoding a single logical qubit \textemdash as the central component of our experimental implementation. The code can be represented graphically, where the physical qubits are the vertices of a $2\times 2$ bicolorable lattice, as shown in~\cref{fig:sc_ls}~(\textbf{Schematic Encoded}) for two initially separate logical qubits labelled $A$ and $B$. Depending on the color, faces are associated with products of either Pauli-$X$ or -$Z$ operators of the adjacent physical qubits. In~\cref{fig:sc_ls}~(\textbf{Schematic Encoded}) for example, the central, orange plaquettes can be associated with operators, $X_1X_2X_3X_4$ and $X_5X_6X_7X_8$. The resulting operators are called \emph{stabilizers} and form a set (group) of operations \textemdash the \emph{stabilizer code} $\mathcal{S}^\textrm{A/B}$ \textemdash under multiplication~\footnote{Note that we choose a negative sign for some stabilizers because this is advantageous for our implementation.},
\begin{equation}\label{eq:sc_stabilizer}
\begin{split}
    \mc{S}^\mr{A}&=\langle S_1^\mr{A},S_2^\mr{A},S_3^\mr{A}\rangle=\langle -Z_1Z_2,-Z_3Z_4,+X_1X_2X_3X_4\rangle,\\
    \mc{S}^\mr{B}&=\langle S_1^\mr{B},S_2^\mr{B},S_3^\mr{B}\rangle=\langle -Z_5Z_6,-Z_7Z_8,+X_5X_6X_7X_8\rangle.
\end{split}
\hspace{5mm}
\raisetag{1.6\normalbaselineskip}
\end{equation}
The logical states $\ket{\psi_\mr{L}^\mathrm{A/B}}$ spanning the respective code spaces for $A$ and $B$ are defined as the simultaneous eigenstates of the stabilizers, i.e., $S_i^\textrm{A/B}\ket{\psi_\mr{L}^\mathrm{A/B}}=\ket{\psi_\mr{L}^\mathrm{A/B}}$, $\forall i\in \{1,2,3\}$. The encoded logical qubits can be associated with logical $X$ and $Z$ operators that anti-commute with each other and commute with all stabilizers. Logical operators are defined up to multiplication with other logical operators, stabilizers and the imaginary unit~$i$. Therefore, the sets of logical operators are defined as
\begin{align}
\begin{split}
    \mathcal{L}^\mathrm{A}&=\langle i, Z^\mathrm{A}_\mathrm{L}, X^\mathrm{A}_\mathrm{L}\rangle/\mathcal{S}^\textrm{A} = \langle i, Z_1Z_3,X_1X_2\rangle/\mathcal{S}^\textrm{A},
    \label{eq:sc_logical}\\
    \mathcal{L}^\mathrm{B}&=\langle i, Z^\mathrm{B}_\mathrm{L}, X^\mathrm{B}_\mathrm{L}\rangle/\mathcal{S}^\textrm{B} = \langle i, Z_5Z_7,X_5X_6\rangle/\mathcal{S}^\textrm{B},
\end{split}
\end{align}
where $\langle P_\mr{L}\rangle/\mathcal{S}$ indicates that logical Pauli operators $P_\mr{L}$ form equivalence classes defined up to multiplication with stabilizers (see~\cref{app:intro_qec} for details). The logical $Y$-operator is determined as $Y_L:=iZ_LX_L$ and we find $Y_\mathrm{L}^\mathrm{A}=Y_1X_2Z_3$ and $Y_\mathrm{L}^\mathrm{B}=Y_5X_6Z_7$. The computational basis states of each logical qubit are then $\ket{0_\mathrm{L}}=\frac{1}{\sqrt{2}}(\ket{0101}+\ket{1010})$ and $\ket{1_\mathrm{L}}=\frac{1}{\sqrt{2}}(\ket{1001}+\ket{0110})$.

Errors can be detected with \emph{error syndromes}, i.e a sign-flip of any code stabilizer. For instance, measuring $S_1^\mathrm{A}$ and obtaining a syndrome $-1$ detects an $X_1$ or $X_2$ error because $\{S_1^\mathrm{A},X_1\}=\{S_1^\mathrm{A},X_2\}=0$. Scaling the surface code is, in theory, as simple as scaling the lattice, see~\cref{app:intro_sc} for more details.\\[-2mm]


\noindent \textbf{Lattice surgery}~\cite{HorsmanFowlerDevittVanMeter2012} is a fault-tolerant\footnote{An operation is called fault-tolerant if errors during the operation can only map to a constant number of physical qubits in the encoding.} protocol for entangling QEC codes which is ideally suited to the geometry of 2D topological codes such as the surface code. This is because LS between topological codes requires only local, few-body interactions. Surface code LS~\cite{HorsmanFowlerDevittVanMeter2012} was introduced as a method to project two surface codes $\mathcal{S}^\mathrm{A}$ and $\mathcal{S}^\mathrm{B}$ with logical operators $X_\mathrm{L}^\mathrm{A},Z_\mathrm{L}^\mathrm{A}$ and $X_\mathrm{L}^\mathrm{B},Z_\mathrm{L}^\mathrm{B}$, respectively, onto joint eigenstates of either $X_\mathrm{L}^\mathrm{A}X_\mathrm{L}^\mathrm{B}$ or $Z_\mathrm{L}^\mathrm{A}Z_\mathrm{L}^\mathrm{B}$, referred to as \emph{rough} and \emph{smooth} LS, respectively~\footnote{Typically, the lattice boundaries of surface codes can be distinguished by their associated stabilizers: $Z$-type stabilizers along the boundary define a \emph{rough} boundary while $X$-type stabilizer define a \emph{smooth} boundary.}. These projections are entangling operations and can be used to construct entangling gates. Here, we proceed by describing rough LS for the minimal $2\times 2$ surface code discussed before and refer to Appendices~\ref{app:intro_ls} and~\ref{app:smooth_ls} for a more general introduction and details.\\[-2mm]

In order to project onto a logical eigenstate of $X_\mathrm{L}^\mathrm{A}X_\mathrm{L}^\mathrm{B}$, we perform a logical joint measurement $M_\mr{XX}^\pm=\mathbb{I}\pm X_\mathrm{L}^\mathrm{A}X_\mathrm{L}^\mathrm{B}$, which can be used to entangle two logical qubits. To achieve this, LS proceeds in two steps: \emph{merging} and \emph{splitting}.
This procedure is illustrated in~\cref{fig:sc_ls}~(\textbf{Schematic Merged}) and~(\textbf{Schematic Split}) for two $2\times 2$ surface codes $\mathcal{S}^\mathrm{A}$ and $\mathcal{S}^\mathrm{B}$. We first \emph{merge} the two separate codes $\mathcal{S}^\mathrm{A},\mathcal{S}^\mathrm{B}$ into a new stabilizer code $\mathcal{S}^\mathrm{M}$ by measuring \emph{merging} stabilizers $S_6^\mathrm{M}=X_3X_5$ and $S_7^\mathrm{M}=X_4X_6$ between the boundaries. These stabilizers commute with all stabilizers of the original codes except $S^\mathrm{A}_2$ and $S^\mathrm{B}_1$, and are chosen such that their joint measurement corresponds to the joint logical measurement $M_\mr{XX}$, i.e., $S_6^\mathrm{M}S_7^\mathrm{M}=X_\mathrm{L}^\mathrm{A}X_\mathrm{L}^\mathrm{B}$. As a result, we obtain the new code by discarding all stabilizers that anti-commute with the merging stabilizers, depicted in~\cref{fig:sc_ls} (\textbf{Schematic Merged}),
\begin{align}\label{eq:sc_merged_rough}
    \mc{S}^\mr{M}&=\langle S_1^\mr{M},S_2^\mr{M},S_3^\mr{M},S_4^\mr{M},S_5^\mr{M}, S_6^\mr{M},S_7^\mr{M}\rangle\nonumber\\
    &=\langle S_1^\mr{A},S_3^\mr{A},S_2^\mr{B},S_3^\mr{B},S_2^\mr{A}S_1^\mr{B},+X_3X_5,+X_4X_6\rangle.
\end{align}
Note that this code already encodes the desired joint eigenstate since $X_\mathrm{L}^\mathrm{A}X_\mathrm{L}^\mathrm{B}$ is included as a stabilizer in the merged code $\mc{S}^\mr{M}$.
In fact, the measurement outcomes $m,m'\in\{0,1\}$ of $S_6^\mr{M},S_7^\mr{M}$, respectively, are random such that $m_1=m+m'$ specifies the eigenvalue associated with $X_\mathrm{L}^\mathrm{A}X_\mathrm{L}^\mathrm{B}$ as $(-1)^\mr{m_1}$.
The merged code is an asymmetric $2\times 4$ surface code encoding a single logical qubit, i.e.,
\begin{align}\label{eq:sc_merged_logic_rough}
    \mc{L}^\mr{M}&=\langle i,Z_\mr{L}^\mr{M},X_\mr{L}^\mr{M}\rangle/\mc{S}^\mr{M}
    =\langle i,Z_\mr{L}^\mr{A}Z_\mr{L}^\mr{B}, X_\mr{L}^\mr{A}\rangle/\mc{S}^\mr{M},
\end{align}
and $Y_\mr{L}^\mr{M}=Y_\mr{L}^\mr{A}Z_\mr{L}^\mr{B}$.
\begin{figure*}[ht!]
\centering
  \includegraphics[width=0.9\textwidth]{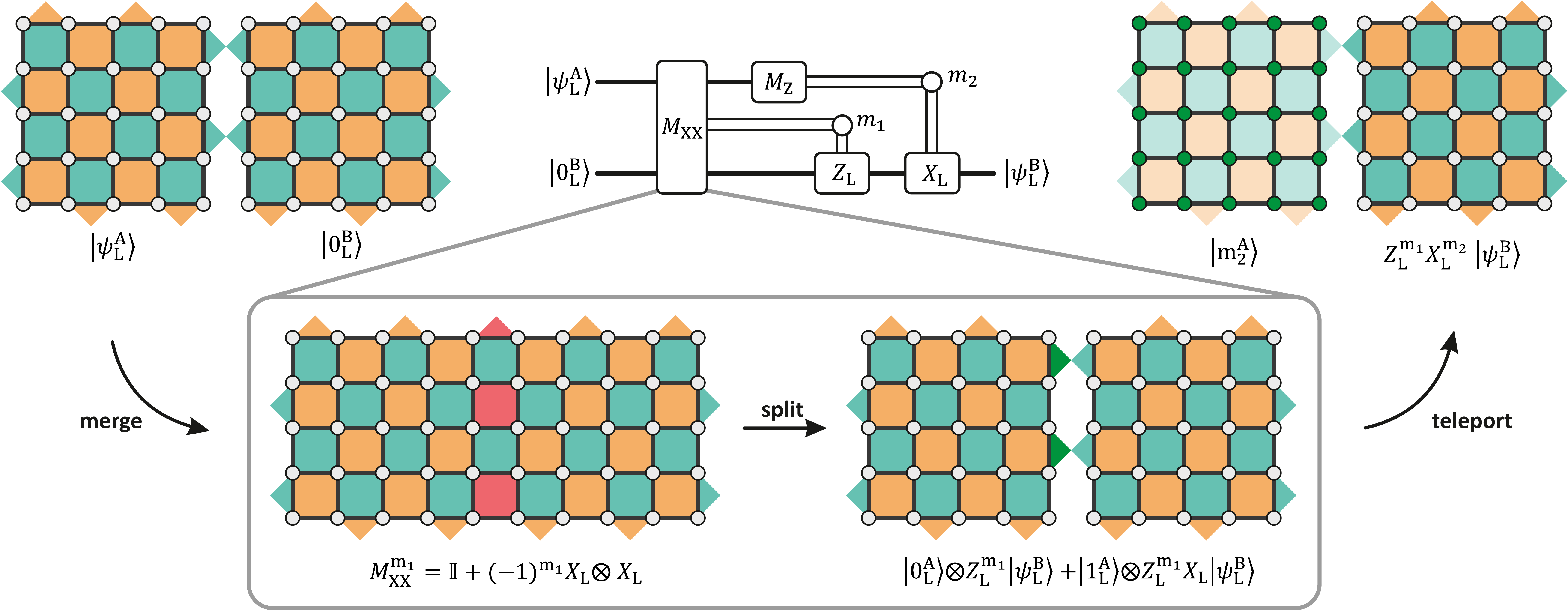}
\caption{\textbf{Surface-code state teleportation with lattice surgery.}
(top-center) Measurement-based scheme to teleport information of an arbitrary logical state $\ket{\psi_\mr{L}}=\alpha\ket{0_\mr{L}}+\beta\ket{0_\mr{L}}$ between two logical qubits using only single-qubit and two-qubit measurements. (left) We start with a logical state $\ket{\psi^\mr{A}_\mr{L}}$ encoded in a $5\times 5$ surface code and an additional logical ancilla in the state $\ket{0^\mr{B}_\mr{L}}$. (bottom) We perform LS, i.e., merging and splitting, to implement a joint measurement $M^\mr{m_1}_\mr{XX}$, where $m_1=0,1$ labels measurement outcomes. The resulting state is entangled. (right) Measuring the initial code in the logical $Z$-basis (i.e., measuring all physical qubits in the $Z$-basis) with measurement outcome $m_2=0,1$ teleports the logical information to the ancilla. Depending on the measurement outcomes $m_1,m_2=0,1$, logical Pauli corrections need to be considered.}
\label{fig:sc_teleportation}
\end{figure*}

With the rough merge we effectively merged the logical $Z$-operators and performed the desired logical joint measurement $M^\pm_\mr{XX}$. Its expectation value $\pm 1$ is given by the product of expectation values of merging stabilizers $S_6^\mathrm{M},S_7^\mathrm{M}$. Now, we must recover the two initial logical qubits while keeping the previously obtained expectation value of $X_\mathrm{L}^\mathrm{A}X_\mathrm{L}^\mathrm{B}$. To this end, we \emph{split} the merged code by measuring $Z$-stabilizers $S^\mathrm{A}_2$ or $S^\mathrm{B}_1$ along the merged boundaries as depicted in~\cref{fig:sc_ls} (\textbf{Schematic Split}). These operators commute with all stabilizers in $\mc{S}^\mr{M}$ that define the separated logical qubits $\mathcal{S}^\mathrm{A},\mathcal{S}^\mathrm{B}$. In particular, the measured stabilizers all commute with $X_\mathrm{L}^\mathrm{A},X_\mathrm{L}^\mathrm{B}$, i.e., the code remains in an eigenstate of $X_\mathrm{L}^\mathrm{A}X_\mathrm{L}^\mathrm{B}$. After splitting, measurement outcomes $m'',m'''\in\{0,1\}$ of stabilizers $S^\mathrm{A}_2, S^\mathrm{B}_1$, respectively, are random but can be tracked as errors. In conclusion, we have effectively performed a logical entangling operation, $M_\mr{XX}^\pm$, which can be used to entangle logical qubits and teleport information.

LS can also be used to realize a measurement-based scheme for logical state teleportation LS~\cite{PoulsenNautrupFriisBriegel2017}. In~\cref{fig:sc_teleportation}, we illustrate this scheme for a logical $M_\mr{XX}$ measurement on two $5\times 5$ surface codes. Note that a similar scheme can be used to teleport information through a logical $M_\mr{ZZ}$ measurement.


\noindent \textbf{Results}. We demonstrate LS in an ion-trap quantum computer, based on atomic $^{40}\mathrm{Ca}^{+}$ ions in a linear Paul trap. Each qubit is encoded in the $\ket{0}=4S_{1/2}(m_j=-1/2)$ and $\ket{1}=3D_{5/2}(m_j=-1/2)$ state of a single ion. Each experiment consists of (i) laser cooling and state preparation, (ii) coherent manipulation of the qubit states, and (iii) readout. (i) For cooling close to the motional ground state, we employ a three-stage process comprising Doppler cooling, polarization gradient cooling~\cite{cirac1992laser,cirac1993laser} and resolved sideband cooling followed by optical pumping into $\ket{0}$. (ii) The qubits are manipulated with a laser at $729$~nm. The available gate set includes single-qubit $Z$ rotations, multi-qubit $X$ and $Y$ rotations and a multi-qubit entangling M{\o}lmer-S{\o}rensen (MS) gate~\cite{sorensen1999quantum}). (iii) Qubit states are read out via electron-shelving. We utilize spectroscopic decoupling to perform operations selectively on a subset of ions by coherently shelving populations from $\ket{0}=4S_{1/2}(m_j=-1/2)$ to $3D_{5/2}(m_j=-3/2)$ and from $\ket{1}=3D_{5/2}(m_j=-1/2)$ to $3D_{5/2}(m_j=+1/2)$. For more details see Ref.~\cite{schindler2013quantum} and~\cref{app:exp_details}.\\[-2mm]

In~\cref{fig:sc_ls}, we demonstrate LS to entangle logical qubits along the rough boundary. Complementary results for smooth lattice surgery are provided in \cref{app:smooth_ls}. We start by encoding the separated logical qubits, each defined by three stabilizers (see~\cref{eq:sc_stabilizer}) and two logical operators (see~\cref{eq:sc_logical}) in~\cref{fig:sc_ls} (\textbf{Encoded}). As a first example, we choose to encode the logical qubits in the state $\ket{0_\mr{L}^\mr{A}0_\mr{L}^\mr{B}}$. We can create encoded states with average stabilizer expectation values of $\langle|S_i|\rangle=0.868(4)$, see~\cref{fig:sc_ls} (\textbf{Code stabilizers Encoded}). We make use of the obtained stabilizer information and post-select our data on states with valid code stabilizers (see~\cref{app:post_selection}), which amounts to discarding those measurements where an error was detected by the code. For the encoded states we infer fidelities of ${\mc{F}(\ket{0_\mr{L}^\mr{A}})=93.8(4)\fps{99.3(2)}}$ and ${\mc{F}(\ket{0_\mr{L}^\mr{B}})=93.4(5)\fps{99.4(2)}}$, where the first value describes the raw fidelity, while the second represents the observed fidelity after post-selection\footnote{This format is used throughout this work to present fidelities of both uncorrected and post-selected data.}. Note that this post-selection introduces a finite survival probability, for details see~\cref{app:survival_probs} and~\cref{app:post_selection}.

Performing LS requires quantum-non-demolition (QND) measurements of stabilizers implemented by series of local and entangling gates (see~\cref{fig:circuit_rough_00}). Considering two merging stabilizers mapped onto ancilla $A1$ and $A2$, we have the possibility to detect one of four possible outcomes $(m,m')=(0,0),(0,1),(1,0),(1,1)$.
In~\cref{app:ancilla_readout}, we present data for all possible outcomes for the chosen input state. For experimental simplicity the following results are for the case $(m,m')=(0,0)$. The merged surface code, as defined in~\cref{eq:sc_merged_rough}, is illustrated in~\cref{fig:sc_ls} (\textbf{Code stabilizers Merged}). The data confirms the merged stabilizers with an average stabilizer expectation value of $\langle|S_i|\rangle=0.669(8)$. Starting from the state $\ket{0^\mr{A}_\mr{L}0^\mr{B}_\mr{L}}$, the merged logical state is a $+1$ eigenstate of the logical $Z_\mr{L}^\mr{M}=Z_\mr{L}^\mr{A}Z_\mr{L}^\mr{B}$ operator, as can be seen in~\cref{fig:sc_ls} (\textbf{Logical operators Merged}). The data reveals a state fidelity of $\mc{F}(\ket{0_\mr{L}^\mr{M}})=86.4(1.0)\fps{97.9(5)}$ after merging.

Now, we split the merged logical qubit along the same boundary by mapping $S_2^\mr{A}$ onto ancilla $A1$ for the case $m''=0$. Thereby we restore the initial code space with an average stabilizer expectation value of ${\langle|S_i|\rangle=0.603(3)}$, shown in \cref{fig:sc_ls}~(\textbf{Code stabilizers Split}). The resulting projective measurement $\mathbb{I}+X_\mr{L}^\mr{A}X_\mr{L}^\mr{B}$ maps the initial product state $\ket{0^\mr{A}_\mr{L}0^\mr{B}_\mr{L}}$ onto a maximally entangled, logical Bell state $\ket{\phi_\mr{L}^+}=\tfrac{1}{\sqrt{2}}\left(\ket{0^\mr{A}_\mr{L}0^\mr{B}_\mr{L}}+\ket{1^\mr{A}_\mr{L}1^\mr{B}_\mr{L}}\right)$.
In order to deduce the fidelity of the generated state with respect to the logical Bell state, we measure the common logical stabilizers $\langle Z_\mr{L}^\mr{A}Z_\mr{L}^\mr{B},X_\mr{L}^\mr{A}X_\mr{L}^\mr{B},-Y_\mr{L}^\mr{A}Y_\mr{L}^\mr{B}\rangle$, obtaining the fidelity
(see, e.g.,~\cite{FriisMartyEtal2018})
\begin{align}
\begin{split}
\mc{F}\left(\ket{\phi_\mr{L}^+}\right)=\tfrac{1}{4}\left(1+\langle Z_\mr{L}^\mr{A}Z_\mr{L}^\mr{B}\rangle+\langle X_\mr{L}^\mr{A}X_\mr{L}^\mr{B}\rangle-\langle Y_\mr{L}^\mr{A}Y_\mr{L}^\mr{B}\rangle\right).\nonumber
\end{split}
\end{align}
In~\cref{fig:sc_ls}~(\textbf{Split}), we present the results for the Bell state generation. From the common stabilizer measurements, we infer a logical Bell state fidelity of ${\mc{F}(\ket{\phi_\mr{L}^+})=58.0(1.6)\fps{75.3(1.6)}}$, where the raw fidelity exceeds the separability limit of $50\,\%$ by $5$ sigma. Imperfect physical gate implementations can be characterized~\cite{erhard2019characterizing} and match our expectations (see~\cref{app:exp_details}). In~\cref{app:additional_meas}, we demonstrate LS for various inputs states in order to generate different maximally entangled Bell states.

\begin{figure}[hbt]
\centering
\includegraphics[width=0.4\textwidth]{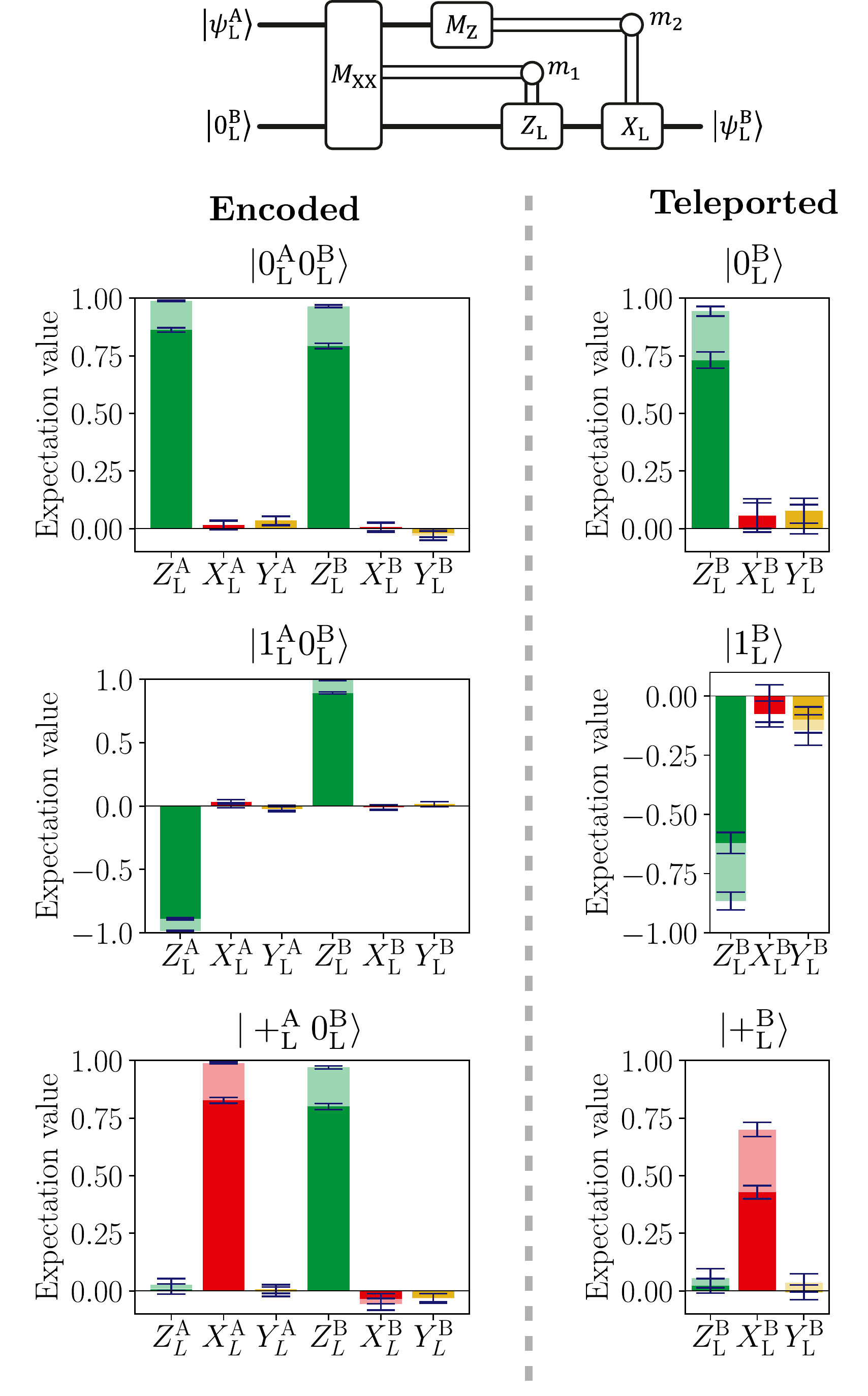}
\vspace*{-4mm}
\caption{\textbf{Teleportation of quantum information via LS.} We prepare the logical qubits $A,B$ in the states $\ket{0^\mr{A}_\mr{L}0^\mr{B}_\mr{L}}$, $\ket{1^\mr{A}_\mr{L}0^\mr{B}_\mr{L}}$, and $\ket{+^\mr{A}_\mr{L}0^\mr{B}_\mr{L}}$, and use LS to teleport the state from logical qubit $A$ to logical qubit $B$. We measure fidelities of the teleported quantum states of $\mc{F}\left(\ket{0^\mr{B}_\mr{L}}\right)=87(2)\fps{97(1)}$  $\mc{F}\left(\ket{1^\mr{B}_\mr{L}}\right)=81(2)\fps{93(2)}$, and $\mc{F}\left(\ket{+^\mr{B}_\mr{L}}\right)=71(1)\fps{85(2)}$.}
\label{fig:teleporation_data}
\end{figure}

LS enables teleporting quantum states from one logical qubit to another (see~\cref{fig:sc_teleportation}), which we demonstrate for the input states $\ket{0^\mr{A}_\mr{L}0^\mr{B}_\mr{L}}$, $\ket{1^\mr{A}_\mr{L}0^\mr{B}_\mr{L}}$, and $\ket{+^\mr{A}_\mr{L}0^\mr{B}_\mr{L}}$. After performing rough LS (i.e., encoding, merging, splitting), we measure logical qubit $A$ in the $Z$-basis and apply a logical $X_\mr{L}$ gate on qubit $B$ if qubit $A$ was found in $\ket{1^\mr{A}_\mr{L}}$ (see~\cref{fig:teleporation_data}). Succeeding the teleportation protocol, we measure logical state fidelities for qubit $B$ of ${\mc{F}\left(\ket{0^\mr{B}_\mr{L}}\right)=87(2)\fps{97(1)}}$,  ${\mc{F}\left(\ket{1^\mr{B}_\mr{L}}\right)=81(2)\fps{93(2)}}$  and ${\mc{F}\left(\ket{+^\mr{B}_\mr{L}}\right)=71(1)\fps{85(2)}}$, given the input states $\ket{0^\mr{A}_\mr{L}0^\mr{B}_\mr{L}}$, $\ket{1^\mr{A}_\mr{L}0^\mr{B}_\mr{L}}$, and $\ket{+^\mr{A}_\mr{L}0^\mr{B}_\mr{L}}$, respectively.

\noindent \textbf{Conclusion}.
We have demonstrated entanglement generation and teleportation via LS between two logical qubits, each encoded in a $4$-qubit surface code, on a $10$-qubit ion trap quantum information processor. We have implemented both the rough and smooth variants of LS~\cite{FowlerMariantoniMartinisCleland2012, PoulsenNautrupFriisBriegel2017}, a technique that is considered~\cite{Litinski2019, GutierrezMuellerBermudez2019} to be key for operating future fault-tolerant quantum computers. For current NISQ-era devices, certification of logical entanglement~\cite{FriisVitaglianoMalikHuber2019} generated via LS can provide means for benchmarking. Besides increasing the numbers of physical and logical qubits, future challenges lie in the implementation of LS between arbitrary topological codes~\cite{PoulsenNautrupFriisBriegel2017} to exploit different features such as transversal gate implementation or high noise tolerance of the respective codes. Lattice surgery can thus function as a fault-tolerant interface between quantum memories and quantum processors.


\vspace*{-2mm}
\bibliographystyle{apsrev4-1fixed_with_article_titles_full_names}
\bibliography{ls}


\vspace*{3mm}
\noindent \textbf{Acknowledgments}.
We acknowledge support from the Austrian Science Fund (FWF) through the SFB BeyondC: F7102. HPN and HJB acknowledge support from the project DK-ALM: W1259-N27. HJB was also supported by the Ministerium f\"ur Wissenschaft, Forschung, und Kunst Baden-W\"urttemberg (AZ:33-7533.-30-10/41/1).
NF acknowledges support from the Austrian Science Fund (FWF) through the project P 31339-N27.
AE, MM, LP, RS, MR, PS, TM and RB acknowledge funding by the U.S. Army Research Office (ARO) through grant no. W911NF-14-1-0103. We also acknowledge funding by the Austrian Research Promotion Agency (FFG) contract 872766, by the EU H2020-FETFLAG-2018-03 under Grant Agreement no. 820495, and by the Office of the Director of National Intelligence (ODNI), Intelligence Advanced Research Projects Activity (IARPA), via the U.S. ARO Grant No. W911NF-16-1-0070. All statements of fact, opinions or conclusions contained herein are those of the authors and should not be construed as representing the official views or policies of IARPA, the ODNI, or the U.S. Government. This project has received funding from the European Union’s Horizon 2020 research and innovation programme under the Marie Skłodowska-Curie grant agreement No. 801110 and the Austrian Federal Ministry of Education, Science and Research (BMBWF). It reflects only the author’s view, the funding agencies are not responsible for any use that may be made of the information it contains. We acknowledge support from the IQI GmbH.\\

\noindent \textbf{Author contributions}.
AE, HPN, PS and NF wrote the manuscript and all authors provided revisions. AE, HPN, PS and NF developed the research based on discussions with HJB, RB and TM. HPN and NF developed the theory. AE and PS performed the experiments. AE, MM, LP, RS, MR, PS, RB and TM contributed to the experimental setup. All authors contributed to discussions of the results and the manuscript.


\hypertarget{sec:appendix}
\appendix

\section*{Appendix: Supplemental Information}

\renewcommand{\thesubsubsection}{A.\Roman{subsection}.\arabic{subsubsection}}
\renewcommand{\thesubsection}{A.\Roman{subsection}}
\renewcommand{\thesection}{}
\setcounter{equation}{0}
\numberwithin{equation}{section}
\setcounter{figure}{0}
\renewcommand{\theequation}{A.\arabic{equation}}
\renewcommand{\thefigure}{A.\arabic{figure}}

In this Appendix/Supplemental Information, we provide additional details on the theoretical background of quantum error correction (QEC) and the performed experiment. In Appendix~\ref{app:intro_qec}, we discuss the stabilizer formalism for QEC, before providing additional details on the surface code and lattice surgery (LS) in Appendices~\ref{app:intro_sc} and~\ref{app:intro_ls}, respectively. In Appendix~\ref{app:smooth_ls} we then explain our experimental realization of smooth LS. Details on the experimental circuits are given in Appendix~\ref{app:exp_details}, followed by information on ancilla readout, survival probabilities for ancilla measurements, and error detection in our setup in Appendices~\ref{app:ancilla_readout},~\ref{app:survival_probs}, and~\ref{app:post_selection}, respectively. Finally, we discuss additional measurements to estimate the logical Bell state fidelities in Appendix~\ref{app:additional_meas}.

\vspace*{-2mm}
\subsection{Stabilizer Quantum Error Correction}\label{app:intro_qec}
\vspace*{-2mm}

In quantum error correction (QEC) our aim is to encode a few logical qubits into many physical qubits such that redundancies can be exploited to detect and correct errors. That is, we replace single-qubit basis states $\ket{0},\ket{1}$ by encoded, logical states $\ket{0_L},\ket{1_L}$ which are made up by many physical qubits. Consider for instance the encoding $\ket{0_L}:=\ket{000}$ and $\ket{1_L}:=\ket{111}$ where we redundantly encoded a logical qubit into three physical (or \emph{data}) qubits. This code can correct for a single-qubit $X$ error. To see this, consider that we observe a state $\ket{100}$. This state is not within the code subspace $\{\ket{0_L},\ket{1_L}\}$ and we have to conclude that either a single qubit error $X_1\ket{0_L}$ or a two-qubit error $X_2X_3\ket{1_L}$ occurred. Under the assumption that a single-qubit error is more likely to occur than two errors on two qubits, we exclude the latter possibility by a majority vote. We can always apply this reasoning to any single-qubit error. In contrast, whenever two-qubit errors occur, our correction would fail and we would effectively introduce a logical error, e.g., $\ket{0_L}\mapsto\ket{1_L}$.

In practice, we won't be able to observe the computational state until the end of a computation. However, we need to be able to perform QEC as described above throughout the computation. Therefore, it is important to note that we can actively observe single-qubit errors in the above code without disturbing the encoded information. This is done by considering so-called code \emph{stabilizers}. These are mutually commuting operators $S_i$ in the three-qubit Pauli group $\mathcal{P}_3$ that map the code subspace to itself while acting as identity on the encoded information, i.e., $S_i\ket{0_L}=\ket{0_L}$, $S_i\ket{1_L}=\ket{1_L}$ and $[S_i,S_j]=0$. Since our code consists of three physical qubits while encoding a single logical qubit, we can expect to find two independent, commuting Pauli operators with this property. Indeed, we find $S_1=Z_1Z_2$ and $S_2=Z_2Z_3$ which can be used to generate a group $\mathcal{S}$ under multiplication, i.e., $\mathcal{S}=\langle S_1,S_2\rangle=\{\mathbb{I},S_1,S_2,S_1S_2\}$. This group is called the \emph{stabilizer group} and contains all stabilizers for this codespace. Since the code subspace is an eigenspace of these operators we can simultaneously measure all stabilizers without disturbing the logical information. Without errors, measuring stabilizers will always result in the same outcome, namely $+1$. However, were an error $X_1$ to occur, the measurement outcome of stabilizer $S_1$, its so-called \emph{syndrome} $s_1$, would change sign since $\{S_1,X_1\}=0$. The only other combination of $X$-errors that could possibly lead to the syndromes $s_1=-1$ and $s_2=+1$ is a two-qubit error $X_2X_3$. Therefore, we end up with the same majority vote as before but without measuring the logical state of the encoded qubit. This is the convenience of the \emph{stabilizer formalism}.

In this formalism, logical operations take a simple form as the \emph{normalizer} $N(\mathcal{S})\subset\mathcal{P}_3$ of the stabilizer group which is the group of operators that leaves the stabilizer group invariant. We are only considering Pauli operators and hence, the normalizer is also the centralizer $C(\mathcal{S})$ of $\mathcal{S}$ which is the group of operators that commutes with all stabilizers. Since this definition includes stabilizers themselves, we define the group of logical operators as a quotient group $\mathcal{L}=N(\mathcal{S})/\mathcal{S}$ such that logical operators form equivalence classes under multiplication with stabilizers. In our case, the equivalence classes are $[\mathbb{I}]_\mathcal{S},[Z_1]_\mathcal{S},[X_1X_2X_3]_\mathcal{S}$, i.e., one for each logical operation.
QEC is done to protect our encoded information from nontrivial logical errors in $\mathcal{L}$. Since we are only considering products of Pauli operators, elements of $\mathcal{L}$ are also just products of Pauli operators. This allows us to infer the maximum number of single-qubit errors our QEC code can tolerate before a logical error occurs, i.e., its \emph{distance} $d$. To see this, consider the nontrivial operator $Z_L\in\mathcal{L}$ and its weight $w(Z_L)$ which is the number of nontrivial terms in the product of Pauli operators. In our example, $Z_L=Z_1$, i.e., its weight is 1 and a single-qubit $Z$-error can cause a logical $Z$-error. In other words, the above code can tolerate no $Z$-errors and its distance is therefore $d=1$. However, w.r.t. logical $X$-operators $[X_1X_2X_3]_\mathcal{S}$, the minimum weight of any logical $X$-operator is 3 such that the code can tolerate three $X$-errors. A code with distance $d$ can generally correct up to $(d-1)/2$ errors and detect up to $d-1$ errors. In our example, our code can correct 1 $X$-error and detect up to $2$.

In summary, QEC in the stabilizer formalism is \emph{active} in the sense that we are required to measure stabilizers and extract syndromes throughout a quantum computation. The syndromes can then be analyzed to determine by a majority vote the errors that have occurred. Logical operators are operators that commute with all stabilizers but are not stabilizers themselves.

\vspace*{-2mm}
\subsection{Surface Code}\label{app:intro_sc}
\vspace*{-2mm}
\begin{figure}[t!]
\centering
\includegraphics[width=0.23\textwidth]{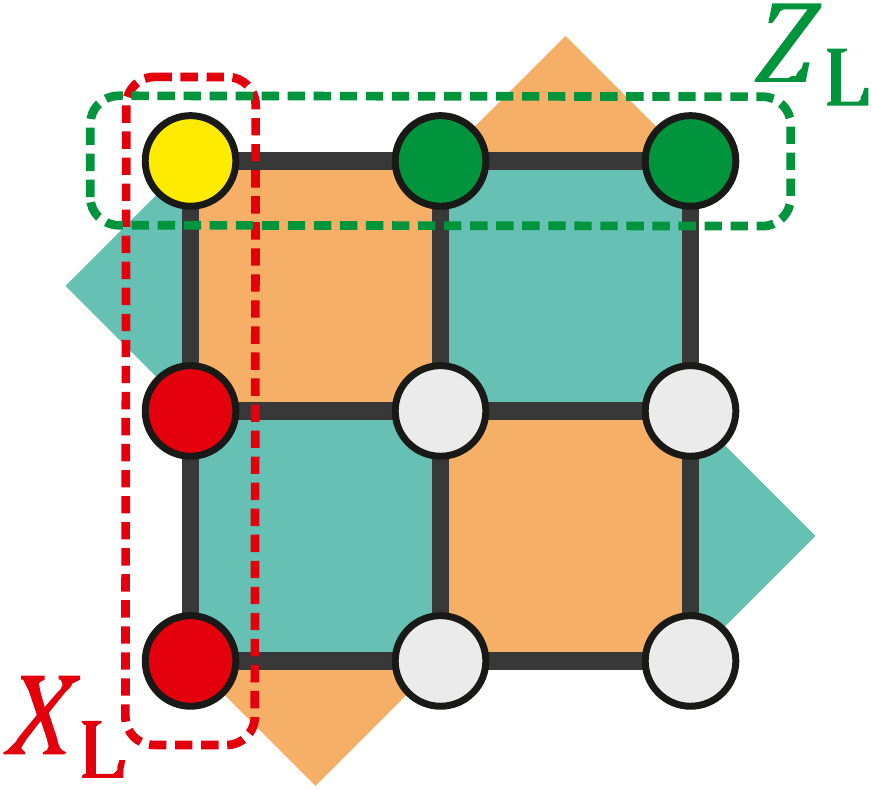}
\vspace*{-2mm}
\caption{
\textbf{Standard surface code of distance 3.}
The standard surface code is defined on a square lattice with (data) qubits located on vertices. Stabilizers are associated with faces and boundaries. Aquamarine faces and boundaries indicate $Z$-type stabilizers as in  Eq.~\eqref{eq:z-stab}. Red faces and boundaries indicate $X$-type stabilizers as in  Eq.~\eqref{eq:x-stab}. The surface code with boundaries encodes a single logical qubit defined by its logical Pauli-$X$ and -$Z$ operators. These operators are defined on strings connecting opposite boundaries of the lattice and act as products of $X$- and $Z$-operators, respectively, along the string. Here two representative logical operators are drawn as products of Pauli-operators within the dashed squares. Red indicates Pauli-$X$ operators and green indicates Pauli-$Z$ operators. The two operators anti-commute at the crossing drawn in yellow.
}
\label{fig:sc}
\end{figure}
Here, we consider a general construction of surface codes in the stabilizer formalism.
Consider $n$ qubits laid out on the vertices $V$ of a bicolored square lattice as displayed in Fig.~\ref{fig:sc}. Let us associate a stabilizer with each colored plaquette $p\in P$ as follows,
\begin{align}
    S^X_{p} = \prod_{v\in\mathcal{N}(p)}X_v\label{eq:x-stab}\\
    S^Z_{p} = \prod_{v\in\mathcal{N}(p)}Z_v\label{eq:z-stab}
\end{align}
where $\mathcal{N}(p)\subset V$ is the set of vertices neighboring a plaquette $p$ and $P$ is the set of faces. $X$-Stabilizers $S^X$ are placed on orange plaquettes while $Z$-Stabilizers $S^Z$ are placed on aquamarine plaquettes. Since neighboring plaquettes always share two vertices, stabilizers commute for all $p\in P$.
For the lattice under consideration, there are $s=n-1$ independent, commuting stabilizers. Therefore, the Hilbert space, which is the simultaneous $+1$ eigenspace of all stabilizer, has $n-s=1$ degree of freedom. This degree of freedom is a qubit since we can define logical $X_L$ and $Z_L$ Pauli-operators. In the case of the surface code, logical operators are products of Pauli-operators connecting opposite boundaries of the lattice. To see this, consider a line drawn on the lattice connecting top and bottom boundaries as indicated by dashed frames in Fig.~\ref{fig:sc}. Placing $X$-operators on vertices crossed by this line, we obtain an operator commuting with all stabilizers but which is not a stabilizer itself. Therefore, this operator corresponds to a logical operator $X_L$. At the same time, we can analogously draw a line for the dual lattice connecting left and right boundary. Placing $Z$-operators along this line, we obtain an operator commuting with all stabilizers but anti-commuting with $X_L$. Therefore, this product of Pauli-$Z$ operators defines the logical $Z$-operator $Z_L$. Note that the shortest line connecting opposite boundaries crosses $3$ vertices. Therefore, the code can tolerate up to three single-qubit errors and has distance $d=3$.

In order to perform QEC, we continuously measure the code stabilizers. Whenever a stabilizer measurement result, i.e., its \emph{syndrome}, changes sign from $+1$  to $-1$, we have detected an error. Assuming that less than $(d-1)/2$ errors have occurred, we can associate with each syndrome a correction procedure which recovers the state of all $+1$ stabilizers from the erroneous state without causing a logical error.

\begin{figure*}[ht!]
  \centering
  \includegraphics[width=0.99\textwidth]{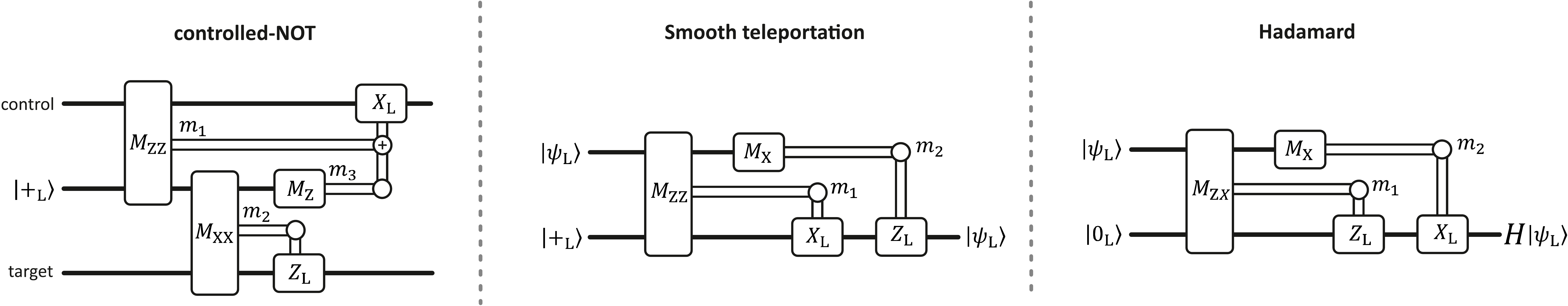}
  \vspace*{-2mm}
\caption{\textbf{Fault-tolerant logic gates with lattice surgery.}
LS enables measurement-based implementations of logic gates and logical state teleportation. LS operations are logical joint measurements of the form $M_{P\tilde{P}}=\mathbb{I}\pm P\tilde{P}$. Thick lines indicate logical qubits in the circuit model and double lines represent classical bits indicating measurement outcomes $m_i=0,1$. Depending on measurement outcomes certain Pauli-corrections need to be applied which are conditioned on the measurement outcomes as $P_L^{m_i}$. $\oplus$ represents an $\operatorname{XOR}$-gate between classical bits. (left) Measurement-based implementation of a logical $\operatorname{CNOT}$-gate between arbitrary control and target qubits requiring an additional ancilla in the $\ket{+_L}$-state. (middle) Measurement-based teleportation protocol for state teleportation between two logical qubits using smooth LS. (right) Measurement-based implementation of a logical Hamadamard gate $H$ based on the teleportation protocol.
}
\label{fig:ls_gates}
\end{figure*}
\vspace*{-2mm}
\subsection{Lattice Surgery}\label{app:intro_ls}
\vspace*{-2mm}
\begin{figure*}[ht!]
\centering
  \includegraphics[width=0.8\textwidth]{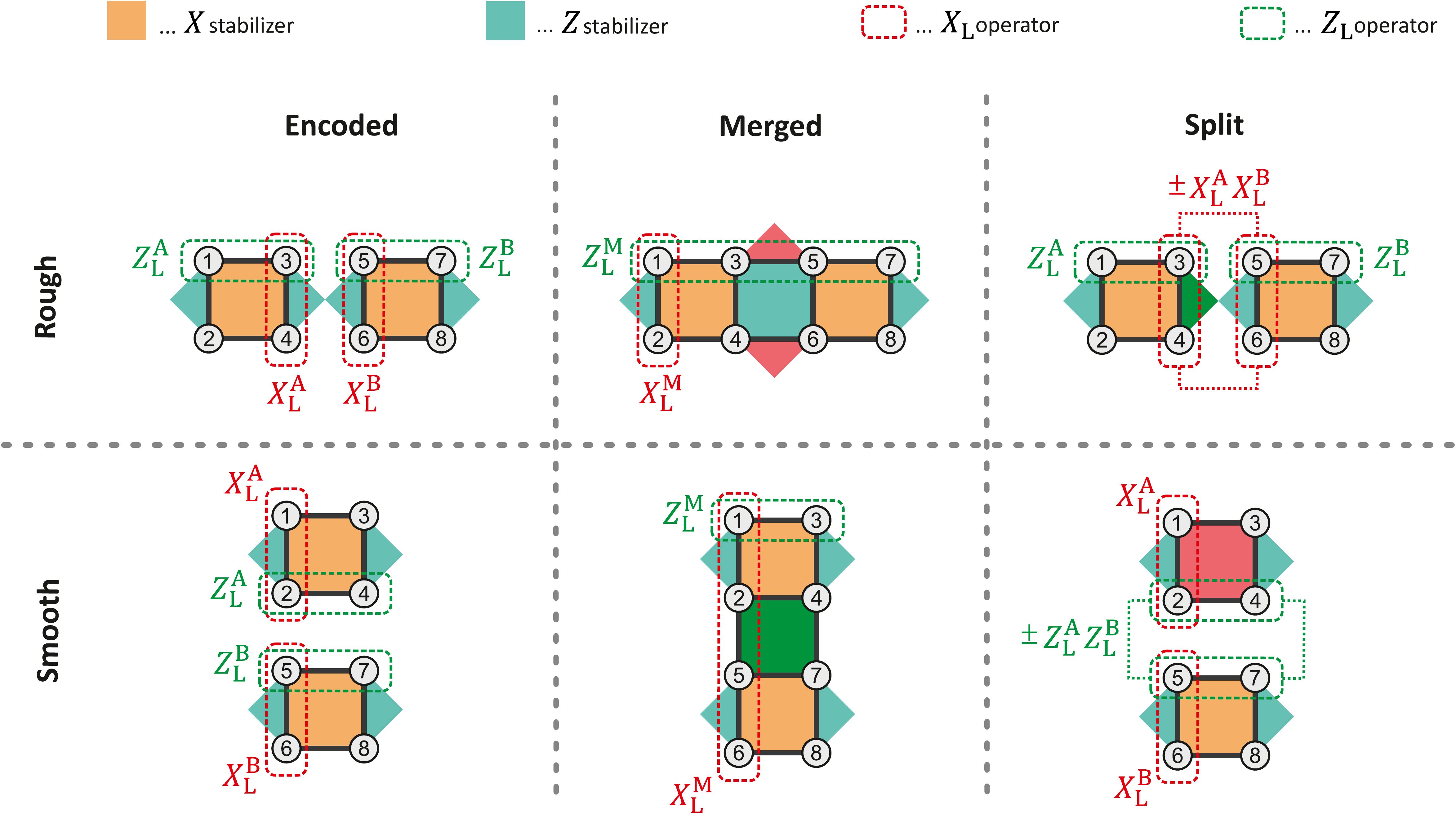}
  \vspace*{-3.5mm}
\caption{\textbf{Surface code lattice surgery.}
Surface code LS between $Z$-type and $X$-type boundaries implementing logical joint measurements $M^{\pm}_\mr{XX}=\mathbb{I}\pm X_\mr{L}^\mr{A}X_\mr{L}^\mr{B}$ (top) and $M^{\pm}_\mr{ZZ}=\mathbb{I}\pm Z_\mr{L}^\mr{A}Z_\mr{L}^\mr{B}$ (bottom), respectively.
\textbf{Encoded:} The two initial surface codes are defined on $2\times 2$ lattices where $X$-stabilizers are associated with orange faces and $Z$-stabilizer with aquamarine faces in accordance with Eq.~(\protect\ref{eq:sc_stabilizer}). Logical operators are products of Pauli operators connecting opposite boundaries as in Eq.~(\protect\ref{eq:sc_logical}). \textbf{Rough Encoded:} The two surface codes are arranged such that they are aligned along their rough (i.e., $Z$-type) boundary. \textbf{Rough Merged:} Treating the two codes as a single (asymmetric) surface code, (merging) stabilizers along the boundaries are measured (indicated in red) such that 
their product
is $X_\mr{L}^\mr{A}X_\mr{L}^\mr{B}$. The merged code encodes a single logical qubit corresponding to the logical Pauli operators $X_\mr{L}^\mr{M},Z_\mr{L}^\mr{M}$. \textbf{Rough Split:} In order to split the merged code while preserving the eigenstate of $X_\mr{L}^\mr{A}X_\mr{L}^\mr{B}$, the boundary stabilizers of the original code are measured (indicated in green). These operators anti-commute with the merging stabilizers and thus project onto the individual codes. Since the boundary operators commute with $X_\mr{L}^\mr{A},X_\mr{L}^\mr{B}$, the resulting state remains an eigenstate of the joint logical operator. \textbf{Smooth} LS: We can project onto a joint eigenstate $Z_\mr{L}^\mr{A}Z_\mr{L}^\mr{B}$ by measuring the $Z$-type merging stabilizer along the smooth boundary analogously to rough LS.
}
\label{fig:app_sc_ls}
\end{figure*}
Here, we consider LS in general as a method to project onto a joint eigenstate of logical Pauli operators.
That is, LS maps two stabilizer QEC codes $\mathcal{S}^{A},\mathcal{S}^{B}$ onto a joint eigenstate $P^{A}_{\mathrm{L}}\otimes \tilde{P}^{B}_{\mathrm{L}}$ of two logical Pauli operators of the codes. This is achieved through a joint measurement $M_{P\tilde{P}}^{\pm}=\mathbb{I}\pm P^{A}_{\mathrm{L}}\otimes \tilde{P}^{B}_{\mathrm{L}}$ which can be implemented fault-tolerantly. In Fig.~\ref{fig:ls_gates}, we illustrate a measurement-based scheme to implement a logical $\operatorname{CNOT}$, Hadamard $H$ and code teleportation using only joint Pauli measurements as described above.

\begin{figure*}[ht!]
\centering
\includegraphics[width=0.95\textwidth]{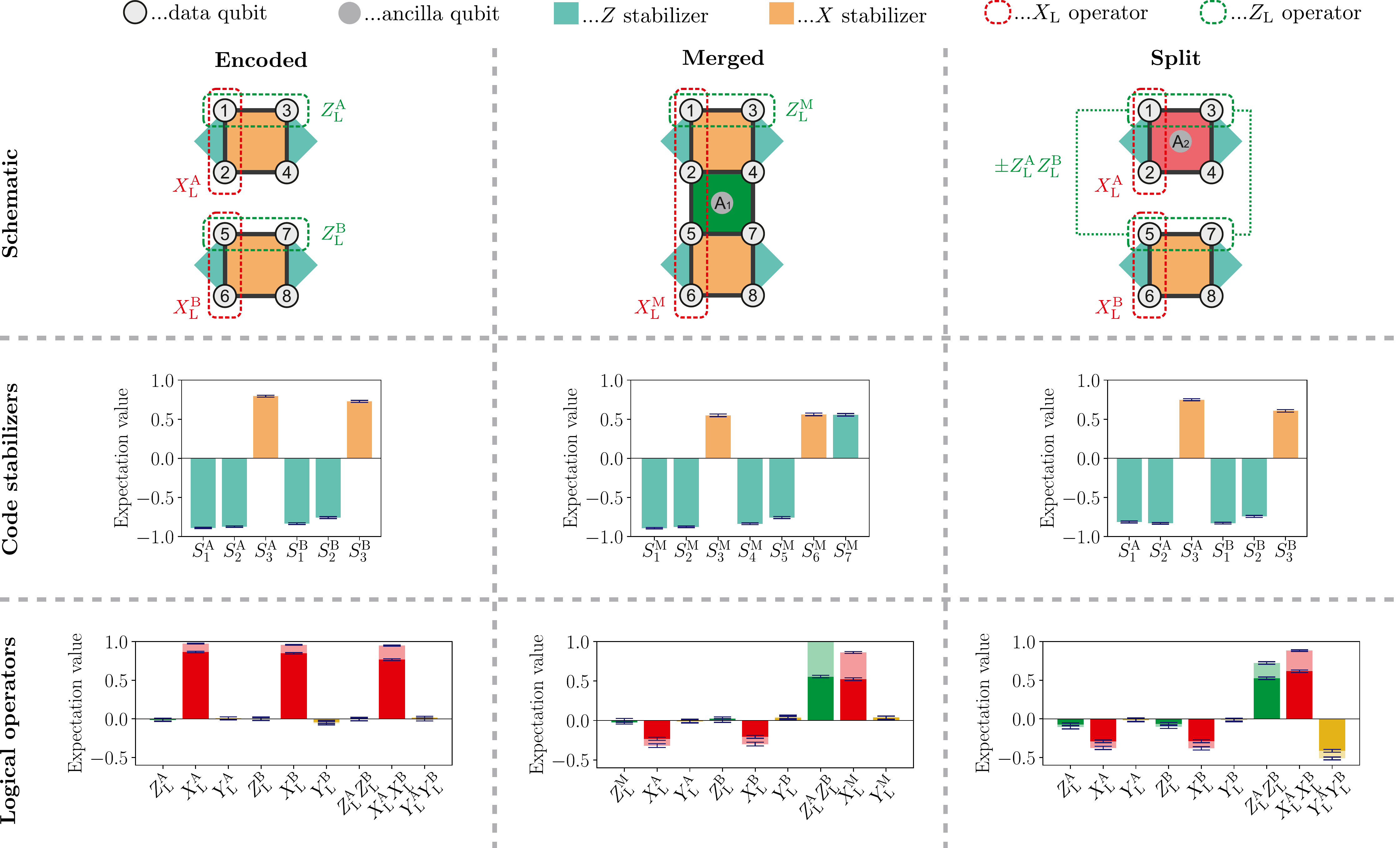}
\vspace*{-3.0mm}
\caption{Bell state generation via lattice surgery along the smooth boundary between two surface code qubits. Post-selected measurements are presented in light colored bars. \textbf{Encoded:} Two logical qubits are encoded with average stabilizer values of $\langle|S_i|\rangle=0.813(4)$. We observe raw and post selected state fidelities for logical qubit A of $\mc{F}(\ket{0_\mr{L}^\mr{A}})=93.3(5)\fps{98.7(2)}$ and for logical qubit B of $\mc{F}(\ket{0_\mr{L}^\mr{B}})=92.4(5)\fps{97.9(3)}$. \textbf{Merged:} The two separated logical qubits are merged into a single logical qubit, whereas the code space is extended in the vertical direction and the new logical operator $X_\mr{L}^\mr{M}=X_\mr{L}^\mr{A}X_\mr{L}^\mr{B}$ is formed. As the data shows the stabilizer $S_7^\mr{M}$ are created. The average stabilizer values and logical state fidelities are $\langle|S_i|\rangle=0.719(5)$ and $\mc{F}(\ket{+_\mr{L}^\mr{M}})=76.2(8)\fps{93.1(6)}$. \textbf{Split:} The single logical qubit is again split into two logical qubits along the same boundary they have been initially merged. We measure the stabilizer $X_1X_2X_3X_4$ to perform the splitting and obtain average stabilizer values of $\langle|S_i|\rangle=0.763(5)$. The fidelity of the generated state with a logical Bell state is $\mc{F}(\ket{\psi_\mr{L}^+})=63.9(2.8)\fps{78.0(2.7)}$.}
\label{fig:smooth_data}
\end{figure*}

LS itself proceeds in two steps: Merging and splitting.
In order to initialize a measurement $M_{P\tilde{P}}^{\pm}$, we first \emph{merge} the two separated codes $\mathcal{S}^{A},\mathcal{S}^{B}$ into a new stabilizer code $\mathcal{S}^{M}$ by projecting onto a joint eigenstate $P^{A}_{\mathrm{L}}\otimes \tilde{P}^{B}_{\mathrm{L}}$. In order for this to be fault-tolerant, we measure so-called \emph{merging} stabilizers $\{S^{M}_i\}_i$ across the boundary such that $\prod_i S^{M}_i=P^{A}_{\mathrm{L}}\otimes \tilde{P}^{B}_{\mathrm{L}}$. This is displayed for the surface code in Fig.~\ref{fig:app_sc_ls} where we consider $\mathcal{S}^{A},\mathcal{S}^{B}$ to be $2\times 2$ surface codes and $P_L=\tilde{P}_{\mathrm{L}}=X_L,Z_L$. Then, the merged code is just a new surface code on an asymmetric lattice and the merging stabilizers are just surface code stabilizers at the interface between the two codes. Stabilizers at the boundary that do not commute with the merging stabilizer are discarded from the stabilizer group and only the product of boundary operators remain since they commute. Notably, the merged code encodes only a single logical qubit and $P^{A}_{\mathrm{L}}\otimes \tilde{P}^{B}_{\mathrm{L}}$ is contained as a stabilizer. That is, this procedure projected onto an eigenstate of $P^{A}_{\mathrm{L}}\otimes \tilde{P}^{B}_{\mathrm{L}}$. The eigenvalue $\pm 1$ is determined by the measurement outcome of the product of merging stabilizers. In order to correct for measurement errors, we need to measure  $\{S^{M}_i\}_i$ $d$ times. That is, measurements can fail and yield a syndrome~$s$ although its expectation value is~$-s$. Such measurement errors can be identified by comparing measurement results at different times.

Now, we want to recover the two initial logical qubits while remaining in an eigenstate of $P^{A}_{\mathrm{L}}\otimes \tilde{P}^{B}_{\mathrm{L}}$. To this end, we \emph{split} the merged code by measuring stabilizers of the separated codes $\mathcal{S}^{A},\mathcal{S}^{B}$ along the aligned boundaries as illustrated in Fig.~\ref{fig:app_sc_ls} (\textbf{Split}). Since these stabilizers anti-commute with merging stabilizers, the set $\{S^{M}_i\}_i$ is discarded from the stabilizer groups and we recover the original two codes. However, since all stabilizers always commute with the logical operators, the resulting state remains an eigenstate of $P^{A}_{\mathrm{L}}\otimes \tilde{P}^{B}_{\mathrm{L}}$. At the end, QEC is required to ensure full fault-tolerance. Surface code LS usually distinguishes \emph{rough} and \emph{smooth} LS when referring to a projection onto a $X_L^{A}X^{B}_{\mathrm{L}}$ or $Z_L^{A}Z^{B}_{\mathrm{L}}$ eigenstates, respectively. However, we do not restrict to smooth and rough LS since a projection onto $Z^{A}_{\mathrm{L}}X^{B}_{\mathrm{L}}$ can be used to generate a logical Hadamard as shown in Fig.~\ref{fig:ls_gates}.


\begin{figure*}[ht!]
\centering
\includegraphics[width=1.0\textwidth]{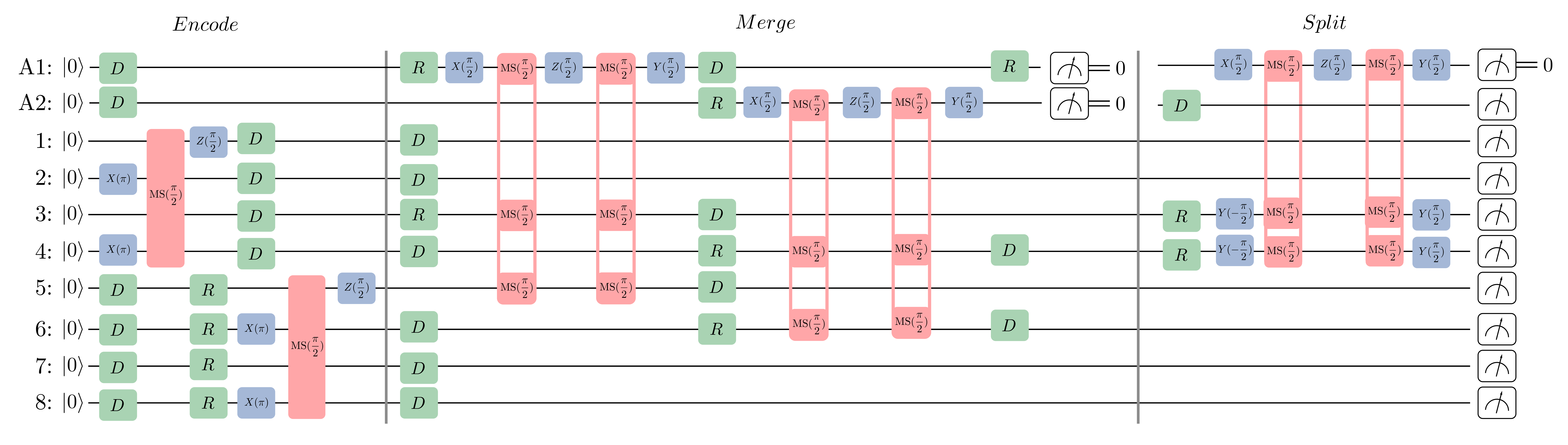}
\vspace*{-7.5mm}
\caption{Circuit diagram for encoding the state $\ket{0_\mr{L}^\mr{A}0_\mr{L}^\mr{B}}$ and doing the merging and the splitting along the rough boundary. Initially all qubits are prepared in the ground state $\ket{0}$. We employ decoupling (D in green) and recoupling (R in green) operations to move qubits in or out of the computational subspace. The local operations are depicted in blue. The multi-qubit entangling MS gates are pictured in red. We implement one in-sequence measurement on the ancilla qubits right after merging the two logical qubits and one measurement of all qubits at the end of the sequence.}
\label{fig:circuit_rough_00}
\end{figure*}

\vspace*{-2mm}
\subsection{Results Smooth Lattice Surgery}\label{app:smooth_ls}
\vspace*{-2mm}

Smooth LS differs from rough LS only in so far that both codes are `rotated' by $90$ degrees before LS. Equivalently, one can understand smooth lattices surgery as merging and splitting along the upper/lower instead of the left/right boundaries, as illustrated in Fig.~\ref{fig:app_sc_ls} (\textbf{Smooth}). In the case of two $2\times 2$ surface codes, measuring the merging stabilizer $S_7^\mr{M}=Z_2Z_4Z_5Z_7$, we obtain a $4\times 2$ asymmetric surface code
\begin{align}\label{eq:sc_merged_smooth}
    \bar{\mc{S}}^\mr{M}&=\langle \bar{S}_1^\mr{M},\bar{S}_2^\mr{M},\bar{S}_3^\mr{M},\bar{S}_4^\mr{M},\bar{S}_5^\mr{M},\bar{S}_6^\mr{M},\bar{S}_7^\mr{M}\rangle
    \nonumber\\
    &
    =\mc{S}^\mr{A}\times \mc{S}^\mr{B}\times \langle +Z_2Z_4Z_5Z_7\rangle.
\end{align}
which can be split by discarding the merging stabilizer.

We present the measured data in~\cref{fig:smooth_data} and obtain a Bell state fidelity of $\mc{F}(\ket{\phi_\mr{L}^+})=63.9(2.8)\fps{78.0(2.7)}$. Further measurement results for various input states can be found in~\cref{app:additional_meas}.


\subsection{Experimental Circuits}
\label{app:exp_details}

As explained in the main text, our available gate set includes local single-qubit $Z$ rotations, multi-qubit $X$ and $Y$ rotations and a multi-qubit entangling M{\o}lmer-S{\o}rensen (MS) gate. The local operations can also be employed on various different Zeeman-transitions in order to spectrocopically decouple and recouple specific qubits from the computational subspace. As an example we present the corresponding circuit diagram for the LS procedure along the rough boundary in~\cref{fig:circuit_rough_00}. In this case we start with encoding the logical qubits in the state $\ket{0_\mathrm{L}}=\frac{1}{\sqrt{2}}(\ket{0101}+\ket{1010})$. Each local operation (depicted in green and blue in~\cref{fig:circuit_rough_00}) consists of a series of multi-qubit $X$ or $Y$ rotations and single-qubit $Z$ rotations. When performing an MS gate, only a subset of qubits is present in the computational subspace (depicted in red in~\cref{fig:circuit_rough_00}). We utilize one 4-qubit MS gate to prepare the state $\ket{0_\mr{L}}$ and we implement two 2-qubit MS gates for the state $\ket{+_\mr{L}}$ plus a number of local gates each. For the stabilizer measurements~\cite{bermudez2017assessing} we employ 6 3-qubit MS gates (rough) and 4 5-qubit MS gates (smooth) as described in~\cref{tab:gates}. To prepare the states $\ket{0_\mr{L}}$ and $\ket{1_\mr{L}}$ we implement the same number of gates but on different qubits. The same applies to the generation of the states $\ket{+_\mr{L}}$ and $\ket{-_\mr{L}}$. If we neglect small multi-qubit gate errors, assume a single-qubit gate fidelity of 99.7\,\% and use the known MS gate fidelities~\cite{erhard2019characterizing} we infer expected Bell state fidelities of $\sim 63\,\%$ (rough) and $\sim 57\,\%$ (smooth). The measured state fidelities of this work of $58(2)\,\%$ (rough) and $64(3)\,\%$ (smooth) reflect the order of magnitude of the expected fidelities very well. We suspect that the largest deviation between the expected and the measured fidelities comes from the error protection while being in a decoherence free subspace in the individual circuits, which was not taken into account.
\begin{table}[hb!]
\caption{Number of gates used for the complete LS circuit (encoding, merging and splitting). We present the number of local 1-qubit gates, local $N$-qubit gates and $N$-qubit MS gates.}
\label{tab:gates}
\begin{tabular}{llllllll}
\hline
Boundary     & Input              & 1-qubit & $N$-qubit & 2-MS & 3-MS & 4-MS & 5-MS \\ \hline
Rough  & $\ket{0_\mr{L}^\mr{A}0_\mr{L}^\mr{B}}$ & 101     & 52        & 0    & 6    & 2    & 0    \\
Rough  & $\ket{+_\mr{L}^\mr{A}0_\mr{L}^\mr{B}}$ & 116     & 60        & 2    & 6    & 1    & 0    \\
Smooth & $\ket{+_\mr{L}^\mr{A}+_\mr{L}^\mr{B}}$ & 121     & 60        & 4    & 0    & 0    & 4    \\ \hline
\end{tabular}
\end{table}


\begin{figure*}[ht!]
\centering
\includegraphics[width=0.8\textwidth]{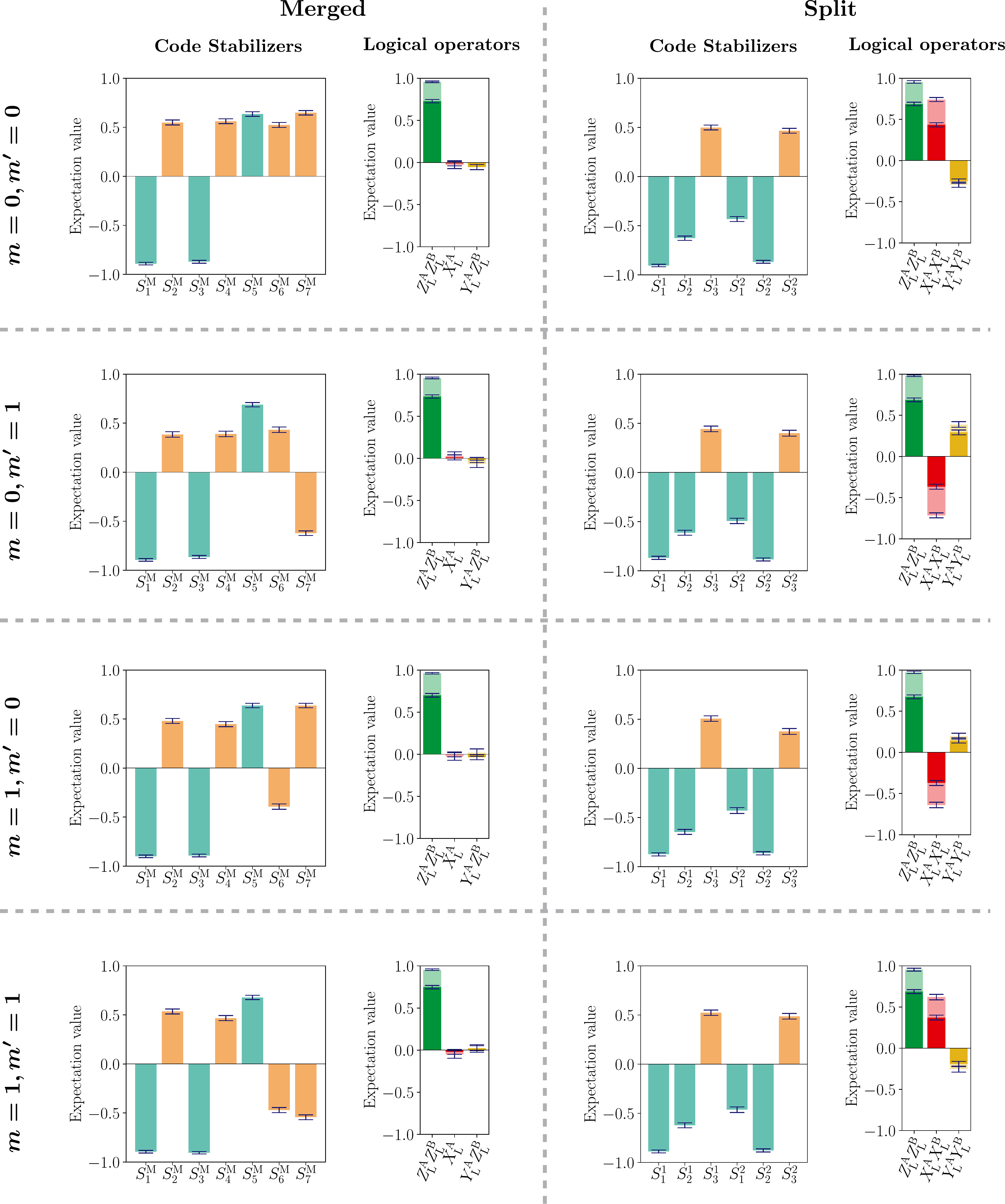}
\caption{Measurements for different outcomes $m,m'$ of the ancilla qubits A1, A2 during rough merging. Starting in the state $\ket{0_\mr{L}^\mr{A}0_\mr{L}^\mr{B}}$, different measurement outcomes are selected by inverting the state of the ancilla qubits right before measurement of the merging stabilizers. The data verifies the expected change of stabilizers $\mc{S}_6^\mr{M}$ and $\mc{S}_7^\mr{M}$ in the merged state and the different resulting Bell states depending on implemented interaction $\mathbb{I}+(-1)^{m+m'}X_\mr{L}^\mr{A}X_\mr{L}^\mr{B}$.}
\label{fig:A1A2_data}
\end{figure*}

\subsection{Ancilla Readout}
\label{app:ancilla_readout}

When merging the two logical qubits, we map the specific stabilizer information onto the ancillas A1 and A2. Subsequently, we measure both ancillas simultaneously. Therefore, we decouple all the data qubits from the computational subspace, and perform a projective measurement only on the ancilla qubits by illuminating the ion string with the 397\,nm laser and collecting the fluorescence light of the ions. Since this is an in-sequence measurement, which means that we will continue with coherent operations after the measurement, we can only use the fast photo-multiplier-tube (PMT) for detection at this stage. The in-sequence measurements come with two difficulties. First, this measurement reveals information about how many ions are bright, but not which ones. Second, the ion chain heats up and the qubits partially leave the computational subspace if the ions scatter 397\,nm light. Without an in-sequence cooling and state preparation technique we cannot do any high-fidelity gate operations after the detection. Hence we can only proceed with the algorithm in the case where both ancilla qubits are found in the dark state $\ket{1}$, where no $397$\,nm photons are scattered. In order to check whether the implemented circuits work faithfully for all possible measurement outcomes, we test the outcome combinations of the ancilla measurements $(m,m')=(0,0),(0,1),(1,0),(1,1)$. In the experiment we only use data where both ancillas where measured to be $m=m'=1$, hence we invert the state of the individual ancilla qubits right before the measurement to investigate any of the four possible outcomes. In~\cref{fig:A1A2_data} we verify the change in the code stabilizers and logical operators for the different outcome combinations of $m$ and $m'$. In our data we observe the expected behaviour, that the stabilizers $\mc{S}_6^\mr{M}$ and $\mc{S}_7^\mr{M}$ change sign depending on the measurement outcomes. The merged state $\ket{0_\mr{L}^\mr{M}}$ is not affected by the measurement. By splitting the merged qubit again we implement the operation $\mathbb{I}+(-1)^{m+m'}X_\mr{L}^\mr{A}X_\mr{L}^\mr{B}$ (see~\cref{fig:A1A2_data}) and the Bell state $\ket{\phi^+}$ or $\ket{\phi^-}$ is generated, depending on the outcomes~$m,m'$.


\vspace*{-2mm}
\subsection{Post-selected stabilizer measurements}
\label{app:survival_probs}
\vspace*{-2mm}

\begin{table}[ht!]
\centering
\caption{Survival probabilities (SP) of stabilizer measurements given in (\%). We present survival probabilities for various input states after merging SP$_i^\mr{M}$ and after splitting SP$_i^\mr{S}$, where $i\in\{X, Y, Z\}$ denotes the respective measurement basis and $\langle\mr{SP}_i\rangle$ describes the average thereof. For each measurement basis and input state we perform 9000 measurements in total.}
\label{tab:surv_probs}
\begin{tabular}{lllllllll}
\hline
Input                    & SP$_Z^\mr{M}$ & SP$_X^\mr{M}$ & SP$_Y^\mr{M}$ & $\langle\mr{SP}^\mr{M}_i\rangle$ & SP$_Z^\mr{S}$ & SP$_X^\mr{S}$ & SP$_Y^\mr{S}$ & $\langle\mr{SP}^\mr{S}_i\rangle$ \\ \hline
$\ket{0_\mr{L}^\mr{A}0_\mr{L}^\mr{B}}$ & 22.5          & 22.7          & 21.9          & 22.4(3)                          & 49.5          & 47.6          & 45.7          & 47.6(19)                         \\
$\ket{0_\mr{L}^\mr{A}1_\mr{L}^\mr{B}}$ & 23.5          & 22.3          & 22.8          & 22.8(5)                          & 47.4          & 49.1          & 47.9          & 48.2(9)                          \\
$\ket{1_\mr{L}^\mr{A}0_\mr{L}^\mr{B}}$ & 22            & 22.7          & 21.6          & 22.1(4)                          & 47            & 48.8          & 47.7          & 47.9(9)                          \\
$\ket{1_\mr{L}^\mr{A}1_\mr{L}^\mr{B}}$ & 21.6          & 22.2          & 21            & 21.6(5)                          & 49.1          & 46.5          & 47.6          & 47.7(13)                         \\ \hline
\end{tabular}
\end{table}

As explained in the section before, we cannot measure all possible ancilla outcomes. When merging the logical qubits along the rough boundary we measure two ancilla qubits, where we only use one of four possible outcomes. In theory this leaves us a survival probability (SP) of 25\,\% of the measurements. If we merge the qubits along the smooth boundary we only use one ancilla and we are left with 50\,\% of the data in theory. In practice we expect this numbers to be lower due to imperfections in the spectroscopic decoupling of the data qubits from the measurement. On the one hand this lowers the SP, on the other hand this increases the fidelity because we already detect certain decoupling errors. When splitting the logical qubit, we map stabilizer information onto one ancilla qubit, which again results in a SP of 50\,\%. This reduction of the survival probabilities can be eliminated by introducing re-cooling and state preparations techniques into the experimental apparatus. We summarize all measured SPs in~\cref{tab:surv_probs}. In addition we get different survival probabilities if we make use of error detection capability of the implemented surface code, which we describe in the next section.


\subsection{Error Detection}
\label{app:post_selection}
\vspace*{-2mm}
The utilized surface code comprising 4 data qubits, is an error detection code with distance (2, 2). Using this code arbitrary single qubit errors can be detected in theory. To detect all single qubit errors, one needs to measure all three code stabilizers of one logical qubit. This requires additional stabilizer measurements on additional ancilla qubits. Since the focus of this work is to show the processing capabilities of LS rather than the error detection capability of the surface code, we leave this to future investigations. But we can use the stabilizers we have in each basis to detect whether or not these stabilizers were correct. For example, if we measure logical qubit A in the $Z$ basis, we check the stabilizers $S_1=Z_1Z_2$ and $S_2=Z_3Z_4$ and detect any single-qubit error on any of the 4 data qubits. In the $X$ basis we detect single qubit errors by checking the stabilizer $S_3=X_1X_2X_3X_4$. If we measure in the $Y$ basis we check the stabilizer $S_2=Z_3Z_4$ and detect any single-qubit error on data qubits 3 and 4, but we do not detect errors on qubit 1 and 2. In general we are not able to detect any 2-qubit or multi-qubit errors. Discarding the measurements with erroneous stabilizer values introduces a finite survival probability, but increases the fidelities significantly compared to the raw fidelities without post selection, as can be observed in all logical stabilizers plots throughout this work. We summarized the uncorrected and post-selected fidelities with the corresponding survival probabilities in~\cref{tab:rough_data}. Since we use multi-qubit MS gates to implement LS, we expect to have multi-qubit errors which cannot be detected and lead to SPs smaller than unity. As can be seen in the data, we detect less single qubit errors in the $Y$ basis and thus we get a higher survival probability SP$_Y$ as expected. Also the 5-qubit MS gate in the smooth merge introduces more multi-qubit errors than the two 3-qubit MS gates in rough merge, which can be observed in the different survival probabilities in the $X$ basis. Using the error detection power of the implemented surface code increases the fidelities by $28(3)\,\%$ with SPs of $64(4)\,\%$ on average.


\vspace*{-2mm}
\subsection{Additional Measurements}
\label{app:additional_meas}
\vspace*{-2mm}

\begin{figure*}[ht!]
\centering
\includegraphics[width=0.8\textwidth]{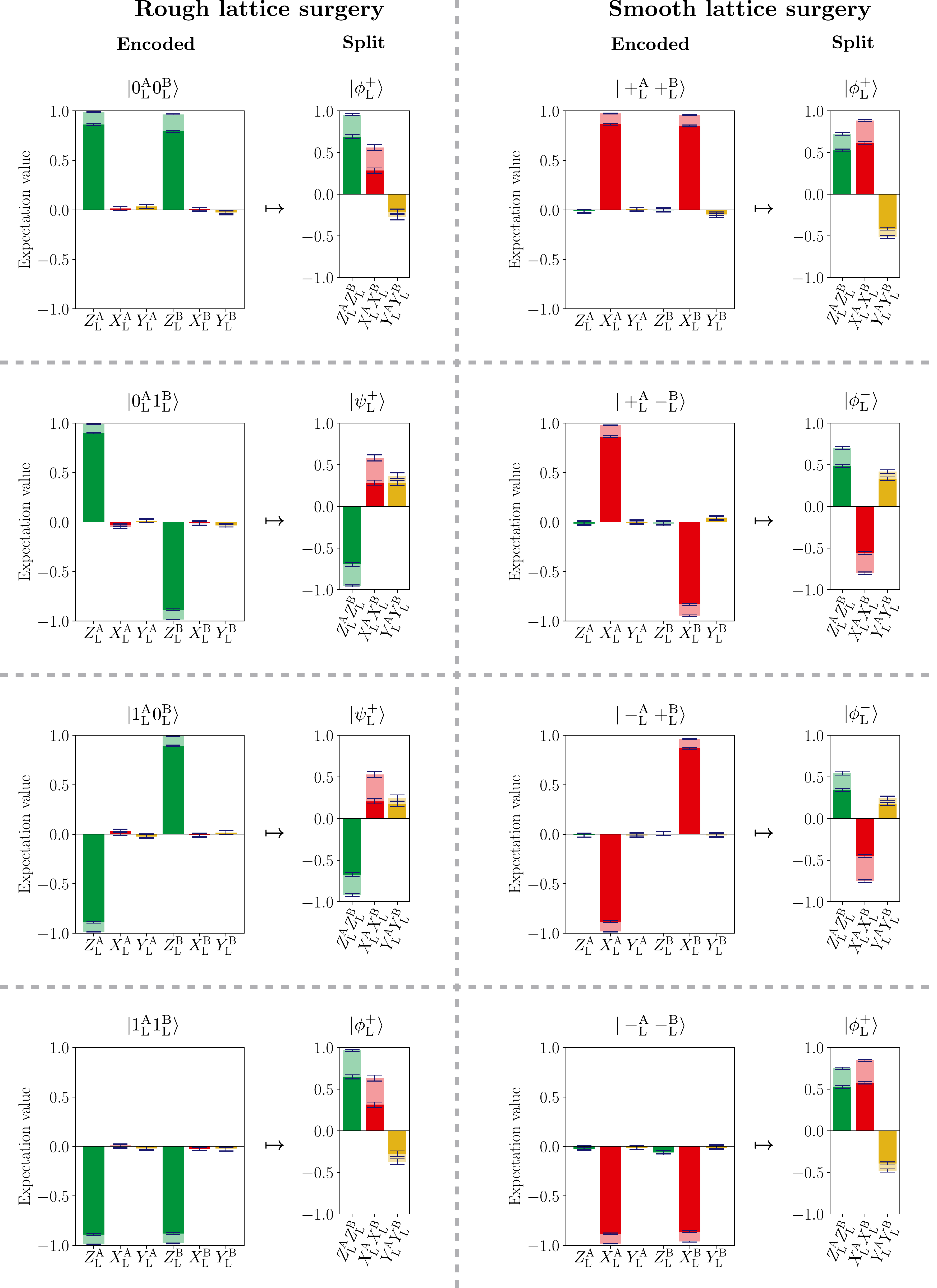}
\caption{Bell state generation with various input states along the rough $\mathbb{I}+X_\mr{L}^\mr{A}X_\mr{L}^\mr{B}$ (left) and the smooth $\mathbb{I}+ Z_\mr{L}^\mr{A}Z_\mr{L}^\mr{B}$ (right) boundary. Detailed information about code stabilizers, fidelities and survival probabilities is given in~\cref{tab:rough_data}. These results should be understood in the same way as~\cref{fig:sc_ls} and~\cref{fig:smooth_data} (\textbf{Logical operators}) before and after LS.}
\label{fig:bell_data}
\end{figure*}

\begin{table*}[ht!]
\centering
\caption{Summary of the Bell state generation experiments for various different input states, along the rough and the smooth boundary. Presented are the individual code stabilizers $\mc{S}_i^A$ and $\mc{S}_i^B$ for qubits $A$ and $B$ as well as the mean of all absolute stabilizer values $\langle|\mc{S}_i|\rangle$. The Bell state fidelities $\mc{F}^\mr{Bell}$ and the post selected fidelities $\mc{F}^\mr{Bell,PS}$ are given in percent (\%). The survival probabilities (SP) after post selection are also displayed in percent (\%). The last column shows the number of taken single shot measurements per basis.}
\label{tab:rough_data}
\begin{tabular}{llllllllllllllll}
\hline
Bound. & Input                                                                                                                                                                                                              & $\mc{S}_1^A$ & $\mc{S}_2^A$ & $\mc{S}_3^A$ & $\mc{S}_1^B$ & $\mc{S}_2^B$ & $\mc{S}_3^B$ & $\langle|\mc{S}_i|\rangle$ & $\mc{F}^\mr{Bell}$ & $\mc{F}^\mr{Bell, PS}$ & SP$_Z$ & SP$_X$ & SP$_Y$ & SP     & Shots \\ \hline
Rough    & $\ket{0_\mr{L}^\mr{A}0_\mr{L}^\mr{B}}$ & -0.87(1)     & -0.69(1)     & 0.30(1)      & -0.49(1)     & -0.84(1)     & 0.42(1)      & 0.603(3)  & 57.9(1.6)          & 75.3(1.6)             & 62     & 49     & 77     & 63(12) & 90000\footnote{Data for plots presented in the main text. The data was taken on a different day with more statistics compared to the rest of data in this table.} \\
Rough    & $\ket{0_\mr{L}^\mr{A}0_\mr{L}^\mr{B}}$                                                                                              & -0.90(1)     & -0.55(3)     & 0.33(3)      & -0.48(3)     & -0.91(1)     & 0.45(3)      & 0.60(1)  & 55.7(5.0)          & 69.7(5.4)             & 62     & 52     & 74     & 63(9)  & 9000\footnote{Data for plots presented in the supplementary text.}  \\
Rough    & $\ket{0_\mr{L}^\mr{A}1_\mr{L}^\mr{B}}$                                                                                                                                                                             & -0.90(1)     & -0.54(3)     & 0.43(3)      & -0.42(3)     & -0.88(2)     & 0.38(3)      & 0.59(1)  & 56.6(4.9)          & 72.7(5.2)             & 59     & 51     & 74     & 61(9)  & 9000  \\
Rough    & $\ket{1_\mr{L}^\mr{A}0_\mr{L}^\mr{B}}$                                                                                                                                                                             & -0.90(1)     & -0.51(3)     & 0.32(3)      & -0.43(3)     & -0.87(2)     & 0.39(3)      & 0.57(1)  & 54.2(5.0)          & 67.3(5.7)             & 59     & 50     & 68     & 59(7)  & 9000  \\
Rough    & $\ket{1_\mr{L}^\mr{A}1_\mr{L}^\mr{B}}$                                                                                                                                                                             & -0.89(1)     & -0.51(3)     & 0.46(3)      & -0.42(3)     & -0.86(2)     & 0.41(3)      & 0.59(1)  & 55.9(5.1)          & 74.3(5.1)             & 58     & 51     & 76     & 62(11) & 9000  \\
Smooth   & $\ket{+_\mr{L}^\mr{A}+_\mr{L}^\mr{B}}$                                                                                                                                                                             & -0.81(1)     & -0.83(1)     & 0.75(1)      & -0.83(1)     & -0.74(1)     & 0.61(2)      & 0.76(1)  & 63.9(2.8)          & 78.0(2.7)             & 70     & 72     & 80     & 74(4)  & 9000  \\
Smooth   & $\ket{+_\mr{L}^\mr{A}-_\mr{L}^\mr{B}}$                                                                                                                                                                             & -0.79(1)     & -0.80(1)     & 0.72(1)      & -0.79(1)     & -0.64(2)     & 0.52(2)      & 0.71(1)  & 59.3(3.2)          & 73.1(3.3)             & 63     & 66     & 73     & 68(4)  & 9000  \\
Smooth   & $\ket{-_\mr{L}^\mr{A}+_\mr{L}^\mr{B}}$                                                                                                                                                                             & -0.58(2)     & -0.65(2)     & 0.63(2)      & -0.75(1)     & -0.62(2)     & 0.49(2)      & 0.62(1)  & 49.3(3.3)          & 63.6(3.8)             & 54     & 62     & 66     & 61(5)  & 9000  \\
Smooth   & $\ket{-_\mr{L}^\mr{A}-_\mr{L}^\mr{B}}$                                                                                                                                                                             & -0.77(1)     & -0.80(1)     & 0.76(1)      & -0.83(1)     & -0.77(1)     & 0.59(2)      & 0.75(1)  & 62.4(2.9)          & 76.8(2.7)             & 69     & 71     & 79     & 73(4)  & 9000  \\ \hline
\end{tabular}
\end{table*}

As explained in the main text, the LS procedure consists of two main parts, first merging two logical qubits into a single logical qubit, and second splitting the logical qubit again into two logical qubits. In total this procedure corresponds to the operation $\mathbb{I}\pm X_\mr{L}^\mr{A}X_\mr{L}^\mr{B}$ (rough) or $\mathbb{I}\pm Z_\mr{L}^\mr{A}Z_\mr{L}^\mr{B}$ (smooth). In our experiment we only implement $\mathbb{I}+X_\mr{L}^\mr{A}X_\mr{L}^\mr{B}$ (rough) and $\mathbb{I}\pm Z_\mr{L}^\mr{A}Z_\mr{L}^\mr{B}$ (smooth) and hence we are able to generate three out of four logical Bell states,
\begin{align}
    \begin{split}
        \ket{\phi_\mr{L}^+}&=\tfrac{1}{\sqrt{2}}\left(\ket{0_\mr{L}^\mr{A}0_\mr{L}^\mr{B}}+\ket{1_\mr{L}^\mr{A}1_\mr{L}^\mr{B}}\right)\\
        \ket{\phi_\mr{L}^-}&=\tfrac{1}{\sqrt{2}}\left(\ket{0_\mr{L}^\mr{A}0_\mr{L}^\mr{B}}-\ket{1_\mr{L}^\mr{A}1_\mr{L}^\mr{B}}\right)\\
        \ket{\psi_\mr{L}^+}&=\tfrac{1}{\sqrt{2}}\left(\ket{0_\mr{L}^\mr{A}1_\mr{L}^\mr{B}}+\ket{1_\mr{L}^\mr{A}0_\mr{L}^\mr{B}}\right).
    \end{split}
\end{align}
The fidelity of the generated state with respect to the logical Bell states can be estimated by measuring the expectation values of the three common stabilizers $\langle Z_\mr{L}^\mr{A}Z_\mr{L}^\mr{B},X_\mr{L}^\mr{A}X_\mr{L}^\mr{B},Y_\mr{L}^\mr{A}Y_\mr{L}^\mr{B}\rangle$ and evaluating the results as follows
\begin{align}
\begin{split}
\mc{F}\left(\ket{\phi_\mr{L}^+}\right)&=\tfrac{1}{4}\left(1+\langle Z_\mr{L}^\mr{A}Z_\mr{L}^\mr{B}\rangle+\langle X_\mr{L}^\mr{A}X_\mr{L}^\mr{B}\rangle-\langle Y_\mr{L}^\mr{A}Y_\mr{L}^\mr{B}\rangle\right),\\
\mc{F}\left(\ket{\phi_\mr{L}^-}\right)&=\tfrac{1}{4}\left(1+\langle Z_\mr{L}^\mr{A}Z_\mr{L}^\mr{B}\rangle-\langle X_\mr{L}^\mr{A}X_\mr{L}^\mr{B}\rangle+\langle Y_\mr{L}^\mr{A}Y_\mr{L}^\mr{B}\rangle\right),\\
\mc{F}\left(\ket{\psi_\mr{L}^+}\right)&=\tfrac{1}{4}\left(1-\langle Z_\mr{L}^\mr{A}Z_\mr{L}^\mr{B}\rangle-\langle X_\mr{L}^\mr{A}X_\mr{L}^\mr{B}\rangle-\langle Y_\mr{L}^\mr{A}Y_\mr{L}^\mr{B}\rangle\right).
\end{split}
\end{align}
In addition to the experiments presented in the main text, we therefore perform LS along the rough boundary for input states $\ket{0_\mr{L}^\mr{A}0_\mr{L}^\mr{B}}$, $\ket{0_\mr{L}^\mr{A}1_\mr{L}^\mr{B}}$, $\ket{1_\mr{L}^\mr{A}0_\mr{L}^\mr{B}}$, and $\ket{1_\mr{L}^\mr{A}1_\mr{L}^\mr{B}}$ and along the smooth boundary for input states $\ket{+_\mr{L}^\mr{A}+_\mr{L}^\mr{B}}$, $\ket{+_\mr{L}^\mr{A}-_\mr{L}^\mr{B}}$, $\ket{-_\mr{L}^\mr{A}+_\mr{L}^\mr{B}}$ and $\ket{-_\mr{L}^\mr{A}-_\mr{L}^\mr{B}}$. The results are presented in~\cref{fig:bell_data} and in~\cref{tab:rough_data}. We observe that the resulting state fidelity $\mc{F}^\mr{Bell}$ for the initial state $\ket{-_\mr{L}^\mr{A}+_\mr{L}^\mr{B}}$ is significantly lower than for the other input states. We suspect that bad calibration is responsible for the decreased fidelity. In general the individual code stabilizers $\mc{S}_i$ have different absolute values, because some stabilizers include two physical qubits, whereas others comprise four physical qubits. Also some physical qubits are more involved in error prone physical gates than others. Hence we do not expect the stabilizer values to be uniformly distributed and we use error propagation to calculate the error of the mean stabilizer values $\langle|\mc{S}_i|\rangle$.
%
%

\end{document}